\documentclass[preprint,12pt]{elsarticle}
\graphicspath{ {./} }
\usepackage{float}
\usepackage{amsmath} % align environment
\restylefloat{figure}
\restylefloat{table}

\begin{document}
\begin{frontmatter}

%\begin{titlepage}
%\begin{center}
%\vspace*{1cm}

%\textbf{ \large \texttt{RanBox}: Anomaly Detection in the Copula Space }

%\vspace{1.5cm}

% Author names and affiliations
%Tommaso Dorigo$^{a,b}$, Martina Fumanelli$^b$, Chiara Maccani$^d$, Marija Mojsovska$^d$, Bruno Scarpa$^c$, Giles C. Strong$^d$ \\

%\hspace{10pt}

%\begin{flushleft}
%\small  
%$^a$ INFN, Sezione di Padova \\
%$^b$ Dipartimento di Scienze Statistiche, Universit\`a di Padova \\
%$^c$ Dipartimento di Scienze Statistiche e Dipartimento di %Matematica, Universit\`a di Padova \\
%$^d$ Dipartimento di Fisica e Astronomia, Universit\`a di Padova \\

%\begin{comment}
%Clearly indicate who will handle correspondence at all stages of refereeing and %publication, also post-publication. Ensure that phone numbers (with country and area %code) are provided in addition to the e-mail address and the complete postal address. %Contact details must be kept up to date by the corresponding author.
%\end{comment}

%\end{flushleft}        
%\end{center}
%\end{titlepage}

\title{\texttt{RanBox}: Anomaly Detection in the Copula Space}

\author[label0]{Tommaso Dorigo}
\ead{tommaso.dorigo@pd.infn.it}

\author[label1]{Martina Fumanelli}
%\ead{second.author@mail.com}

\author[label2]{Chiara Maccani}
%\ead{author@mail.com}

\author[label2]{Marija Mojsovska}
%\ead{author@mail.com}

\author[label2]{Giles C. Strong}

\author[label1]{Bruno Scarpa}
\ead{scarpa@stat.unipd.it}

\address[label0]{INFN - Sezione di Padova, Via F. Marzolo 8, 35131 Padova, Italy}
\address[label1]{Dipartimento di Scienze Statistiche, Universit\`a di Padova, Via C. Battisti 241, 35131 Padova, Italy}
\address[label2]{Dipartimento di Fisica e Astronomia ``G.Galilei'', Universit\`a di Padova, Via F. Marzolo 8, 35131 Padova, Italy}

\begin{abstract}
The unsupervised search for overdense regions in high-dimensional feature spaces, where locally high population densities may be associated with anomalous contaminations to an otherwise more uniform population, is of relevance to applications ranging from fundamental research to industrial use cases.  Motivated by the specific needs of searches for new phenomena in particle collisions, we propose a novel approach that targets signals of interest populating compact regions of the feature space. The method consists in a systematic scan of subspaces of a standardized copula of the feature space, where the minimum $p$-value of a hypothesis test of local uniformity is sought by gradient descent. We characterize the performance of the proposed algorithm and show its effectiveness in several experimental situations.
% GS: This is the only time that "loss function" and "gradient descent" are mentioned in the paper, and as far as I understand, the optimisation isn't via gradient descent... --> yep removed
\end{abstract}


\end{frontmatter}

\clearpage
%%%%%%%%%%%%%%%%%%%%%%%%%%%%%%%%%%%%%%%%%%%
\section{Introduction \label{introduction}}

\subsection{Searches for new phenomena in collider physics}

\noindent
The word {\em anomaly} has its roots in the ancient Greek word $\alpha \nu \grave{\omega} \mu \alpha \lambda o s$; in common parlance, anomalous means ``different, peculiar, or not easily classified''~\cite{merriam-webster}. Consistently with the common usage of the word, in Statistics an anomalous datum is one which does not conform to the others in a set, because its observable features $\theta$ single it out as unlikely to have been sampled from the probability density function\footnote { The subscript $b$ denotes it as the distribution of the ``background'', which is the name commonly associated to the non-anomalous component of unknown data.} $p_b(\theta)$ which the rest of the data conform to. 

In the context of searches for new phenomena in high-energy particle physics (HEP), the identification of anomalies is of great interest. At the Large Hadron Collider (LHC), which produces the most energetic subnuclear reactions ever achieved in a laboratory, the CMS~\cite{CMS} and ATLAS~\cite{ATLAS} experiments compare the observable features of the final state of proton-proton collisions to extremely precise predictions yielded by the Standard Model (SM) of particle physics~\cite{gsw}, in search for signals of new physics that the SM does not account for. Each collision can be typically summarized, through a complex reconstruction of tens of millions of digitally recorded signals, into few dozens of high-level features. The comparison of the distribution of those features with the ones expected from SM processes yields sensitivity to new physics phenomena. Given the large dimensionality of the problem, the typical approach followed by CMS and ATLAS in their searches is supervised classification: Monte Carlo simulations of both SM processes and hypothetical new physics phenomena inform the training of a classifier, whose output enables inference on the existence of new physics.

The above {\em modus operandi} rests on two pillars: reliance on an extremely precise model of $p_b(\theta)$, from which SM processes are sampled, and availability of theoretical models predicting the possible distribution of the density function $p_s(\theta)$ of signal events. It should be clear that the conditions on which those two pillars are based are difficult to satisfy in practice.

The SM, while superbly tested through decades of experimental probing~\cite{ewwg}, is subjected to uncertainties arising from imperfect knowledge of its underlying parameters, as well as from the purely empirical description of some of the fundamental ingredients playing a role in the particle collisions --{\em e.g.}, the parton distribution functions of the colliding protons, which determine the relative probability of different processes and their energy release. The resulting systematic uncertainties are especially impactful in those poorly-investigated regions of phase space which, thanks to the LHC's unprecedented reach, are probed by experimental data for the first time, and which are consequently the most likely hiding place of new processes. In addition to systematic uncertainties, new physics models suffer from the limited range of possibilities that they explore. New physics may manifest itself in ways that theoreticians have not yet ventured to speculate; if the resulting high-level features of signal events are not striking enough, or if they do not result in conspicuous modifications of some of the marginals of $p_b(\theta)$, they may be overlooked.

For the above reasons it appears necessary to plan for a systematic, unsupervised exploration of the feature space of LHC collider data. This task must be pursued with a diversified weaponry, such that sensitivity to a range as wide as possible of new physics phenomena is achieved. In this document we describe a contribution in that direction. The algorithm we developed, called \texttt{RanBox}, is designed to exploit and fit to the characteristics of collider data: in particular, the wide range of values taken by $p_b(\theta)$, which calls for a vigorous standardization procedure at preprocessing stage; its local smoothness in the feature space; and the typically limited phase space where a signal density $p_s(\theta)$ may contribute in an observable way to recorded data. We however stress that, while conceived with the specific task of achieving sensitivity to new physics phenomena at the LHC, and designed around it, the developed algorithm may prove useful to a wide range of applications that share the need for sensitivity to localized overdensities in the feature space. 

The plan of this document is as follows. In the remainder of this Section we briefly formalize the problem we wish to solve. In Sec.~\ref{s:algorithmdescription} we describe the \texttt{RanBox} algorithm and its variant \texttt{RanBoxIter}. In Sec.~\ref{s:synthetic} we demonstrate the performance of the algorithm on a synthetic dataset of simplified characteristics. We proceed to exemplify the results of the application of \texttt{RanBox} and \texttt{RanBoxIter} on several publicly available datasets in Sec.~\ref{s:experiments}. In Sec.~\ref{s:related} we mention other studies that address the issue of unsupervised or semi-supervised searches of new phenomena in particle collisions, and their relation to \texttt{RanBox}. We offer some concluding remarks in Sec.~\ref{s:concludingremarks}.

%%%%%%%%%%%%%%%%%%%%%%%%%%%%%%%%%%%%%%%%%%%%%%%%%%%%%%%
\subsection {Problem Statement \label{s:problemstatement}}

\noindent
We consider a set of $N$ data examples ${x} \in \cal{S} \subseteq \mathbf{R}^D$ sampled from a unknown multivariate density function $p(x)$. In general, $p(x)$ can be written as the sum of a background component $p_b(x)$ and a possible signal contamination $p_s(x)$,

\begin{equation}
    p(x) = (1-f_s) p_b(x) + f_s p_s(x)
\end{equation}

\noindent
where $f_s$ is the signal fraction. An anomaly detection problem may be defined as one of finding a localized region of the feature space $\cal{S}$ that contains a density of data examples significantly higher than that of its surroundings, as defined by some suitable metric. This problem may be cast as a semi-supervised or a unsupervised one, depending on whether the (by definition) non-anomalous density of the background component is assumed known or not; in both cases, a central issue is how to retain sensitivity to a wide variety of anomalous contaminations, which may produce distortions of the density in a subset of the ${\cal{D}}$ features.

%%%%%%%%%%%%%%%%%%%%%%%%%%%%%%%%%%%%%%%%%
\subsection {The idea of \texttt{RanBox}}

\noindent
In this work we consider the unsupervised version of the problem, which offers the benefit of avoidance of any model-related uncertainties, and we focus on new physics signals that characteristically produce localized, compact variations in the overall density of the feature space.

We wish to construct an algorithm that searches the feature space $\cal{S}$ by considering a ``box'', {\em i.e.}, a multidimensional interval constructed in a subspace of $\cal{S}$. The random nature of the box lays not only in the endpoints $x_i^{min}$, $x_i^{max}$ of its intervals in each marginal, $x_i \in [x_i^{min},x_i^{max}]$, but also in the involved subspace $\cal{S'} \subseteq \mathbf{R}^{D'}$ of $\cal{S}$ described by a subset of the $x_i$. Alternatively, one may think of the box as having restricted intervals in only a subset $\cal{D}-\cal{D'}$ of the dimensions of $\cal{S}$. If we consider for the time being the case $f_s=0$ and $N$ data points in $\cal{S}$ sampled from a multi-dimensional uniform density $p_b(x)= {\cal{U}}(x)$, such a box will contain a predictable fraction of the total data: given the box volume $V_{box}$ and the total volume $V$ of the feature space  ${\cal{S}}$, the expectation value of the number of events in the box is $N_{exp} = N V_{box}/V$. Conversely, if $f_s>0$, the observed number of events captured within the box boundaries $N_{obs}$ may yield an estimate of the density of the total sampling distribution in the corresponding region of $\cal{S}$, contributed by both $p_b(x)$ and $p_s(x)$: \par

%\large
\begin{equation}
\hat{p}(x) = \frac{N_{obs}}{V N_{exp}} = \frac{N_{obs}}{N V_{box}}.    
\label{eq:estimate}
\end{equation}
%\normalsize

\noindent
The above estimate may be used to construct a test statistic sensitive to an anomalous local overdensity of the data; {\em e.g.}, one may simply define the test statistic to equal the estimated excess of events in the box, $N_{obs}-N_{exp}$, or a significance measure of its non-null value. The maximization of such a test statistic will be appropriate for searches of anomalies that preferentially populate well-confined regions of the feature space, such as those of interest in collider searches for new physics, but also relevant to other branches of science as, {\em e.g.}, astrophysical observations, or industrial applications such as process control, fraud detection, or spam filtering. Conversely, we expect little sensitivity to multi-modal signals, and (by construction) no sensitivity to broad deformations of a nearly-uniform background distribution $p_b(x)$. The locality of the signal to be detected, however, is the only assumption we allow ourselves to take in the construction of our anomaly detection procedure. The assumption of uniformity on which the estimate in Eq.~\ref{eq:estimate} is based can be loosened if we work in the copula space, as discussed more in detail in Sec.~\ref{s:algorithmdescription}.

%%%%%%%%%%%%%%%%%%%%%%%%%%%%%%%%%%%%%%%%%%%%%%%%%%%%%%%%%%%%%%
\section{Algorithm Description \label{s:algorithmdescription}}

%%%%%%%%%%%%%%%%%%%%%%%%%%%%%%%%%%%%%
\subsection {Starting considerations}

\begin{figure}[h!]
\begin{center}
\includegraphics[width=12cm]{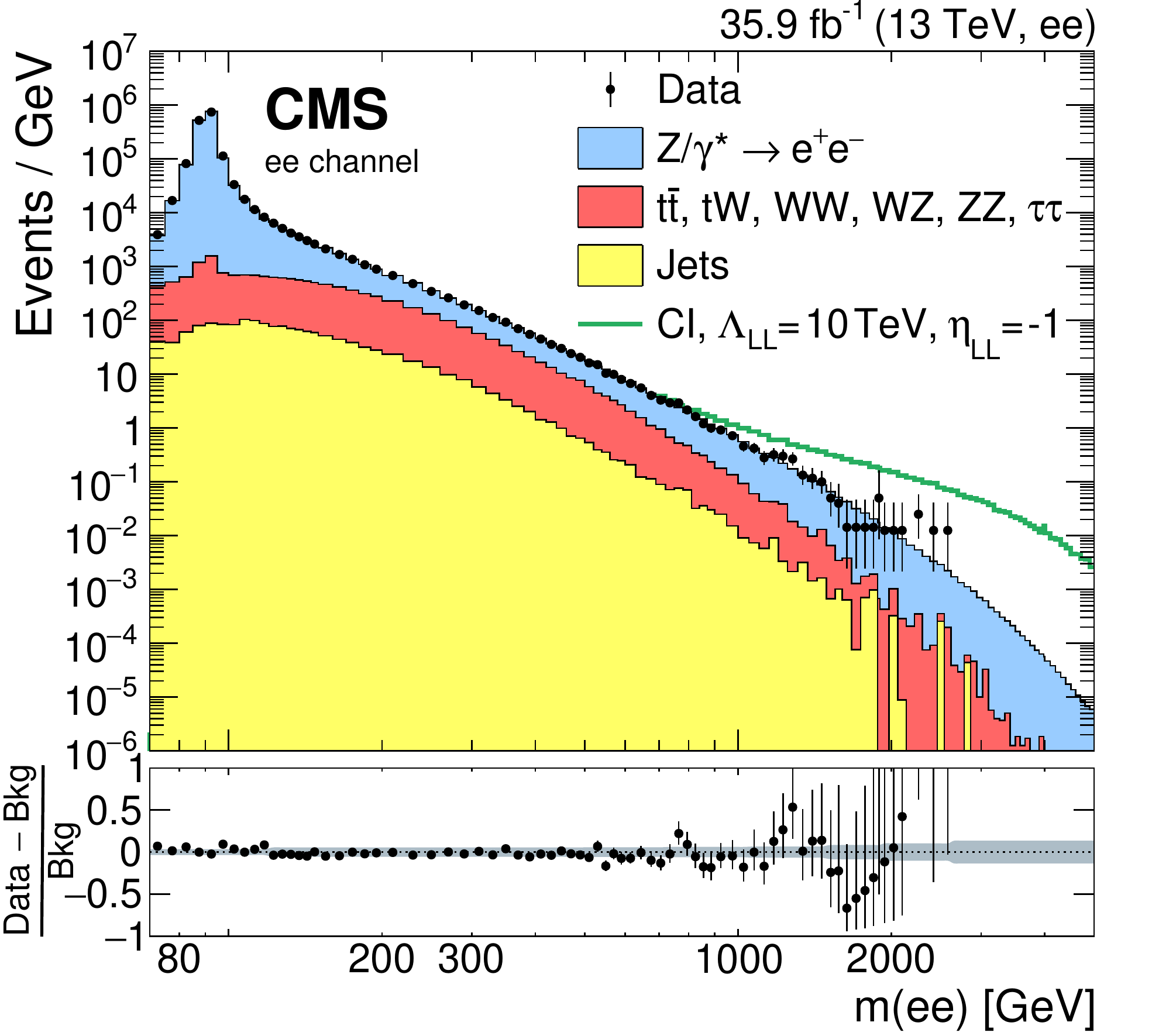}
\caption{\small Distribution of the invariant mass of candidate electron-positron pairs observed by the CMS experiment in 36 $fb^{-1}$ of Run 2 LHC collisions~\cite{cmsdielectrons}. The data show a variation in density by several orders of magnitude as a function of mass. The cited reference reports on searches for a new physics contribution involving contact interactions, which could contribute to the distribution at its high-end tail (green curve). }
\label{f:expdistr}
\end{center}
\end{figure}

\noindent
The multitude of subnuclear particles resulting from proton-proton collisions recorded by LHC experiments, which we take as our target application in the construction of the algorithm, yield tens of millions of electronic signals in the detectors. This large body of information is summarized by a process called ``event reconstruction'' through the extraction of several tens of high-level features that describe the measurement of energy and direction of all observed particles ({\em e.g.}, energetic electrons or muons) or sets of particles (hadronic jets)~\footnote{ In HEP it is thus customary to call {\em events} the observed data examples, and we will stick to that convention in this work.}. Even if we focus on specific interesting subsets of the available data, any energy-related feature of the observed particles will show a highly dis-uniform distribution, with a peak at low values and long tails extending to higher energy (see, {\em e.g.}, Fig.~\ref{f:expdistr}). The variation in density between those peaks and tails may amount to orders of magnitude, and is due to the corresponding large variation in the probability that the collision is originated by quarks or gluons carrying a low or a high fraction of their parent's total momentum. 

\noindent
Because of the above, it seems natural to proceed by first pre-processing the data with an integral transform of all the features, such that each marginal becomes uniform by construction. The algorithm will then work in the copula space, examining the data structure with a metric unaffected, at least to first order, by the original strong density variations in the feature space.

%%%%%%%%%%%%%%%%%%%%%%%%%%%%%%%%%%%%%%%%%%%%%%%%%%%%%%
\subsection {Data preprocessing \label{s:preprocessing}}

\noindent
The probability integral transform of a function $f(x)$ is defined by setting \par

\begin{equation}
F(x) = \int_{-\infty}^{x}f(t)dt, 
\end{equation}

\noindent
which is such that $y=F(x)$ is uniform in $[0,1]$: \par

% \begin{equation}
% \label{eq:standardize1}
% F_y(y) = P(Y \leq y)
% \end{equation}
% \begin{equation}
%       = P(F_X(X) \leq y) 
% \end{equation}
% \begin{equation}
%       = P(X \leq F_x^{-1}(y))  
% \end{equation}
% \begin{equation}
%       = F_X(F_X^{-1}(y)) = y .
% \label{eq:standardize2}       
% \end{equation}
% GS: replaced with align environment below:

\begin{align}
    F_y(y) &= P(Y \leq y) \notag\\
   &= P(F_X(X) \leq y) \notag\\
   &= P(X \leq F_x^{-1}(y)) \notag\\
   &= F_X(F_X^{-1}(y)) = y.
   \label{eq:standardize1}
\end{align}

\noindent 
Once each of the variables of the feature space ${x_i}$ is transformed as above into the corresponding one in the set ${y_i}$, information once contained in the interdependence of the $x_i$ is retained in the copula, which is the joint distribution function of variables with uniform marginals (Sklar's theorem)~\cite{sklar}. The advantage of the transformation is evident: a search for overdensities in the space spanned by ${y_i}$ will not be spoiled by uneven marginals, and will correctly concentrate on the regions of space which are dense because of interdependence of the features. An additional bonus of working with the ${y_i}$ variable basis is that the feature space is now a unit hypercube, with volume $V=1$. 

%%%%%%%%%%%%%%%%%%%%%%%%%%%%%%%%%%%%%
\subsection{Dimensionality reduction}

\noindent
The dreaded ``curse of dimensionality''~\cite{bellman} affects any search in high-dimensional spaces populated by sparse data. In the typical applications considered in this work, the total data size $N$ lays in the few thousands to few hundreds of thousands range; consequently, an investigation of subspaces ${\cal{S'}}$ of the feature space ${\cal{S}}$ quickly becomes meaningless as their dimensionality grows larger than about ${\cal{D'}}=12-15$, when Poisson fluctuations prevent any reasonable multi-dimensional density estimate. 

An additional optional preprocessing step, which may prove useful to reduce the dimensionality in cases when $\cal{D}$ is larger than a few tens, is the application of Principal Component Analysis (PCA) to the feature space. PCA essentially consists in fitting a hyper-ellipsoid to the data, and remapping the feature space in a space spanned by the principal axes of the ellipsoid. One may then use the principal components, which are those on which the data exhibit the largest variance, and ignore the last few in the ordered list of components, which are likely to contain the least information. PCA can be useful for \texttt{RanBox} in cases when the search for subspaces of limited dimensionality $\cal{D'}$ of the feature space proves impractical because of the large binomial coefficient ${{\cal{D}}\choose{\cal{D'}}}$% \binomial{D'}{D}$
, which makes the exploration of a meaningful fraction of the possible $\cal{D'}$-dimensional subspaces too CPU-intensive. However, in our investigations we have found that PCA is generally liable to reduce the power of the search for overdense regions of the feature space when the data are composed of a large background component and a small signal contamination to which we wish to be sensitive. The typical reason of this effect is connected with the fact that a variable which exhibits little variance on the majority of the data, and is thus discarded by PCA, may still be very distinctive for a small signal. An example of this behaviour is given in Sec.~\ref{s:experiments_miniboone}.

A viable alternative to reduce the dimensionality of the problem, which may facilitate the identification of small signals, is to exploit the correlation matrix of the features, by removing features which add little information. This is an attractive option when searching for small anomalous components in a background-rich dataset: by identifying and removing variables that are highly correlated with others on the majority component of the data, we reduce the possibility that such correlations affect negatively the chance of the algorithm to identify localized overdensities genuinely due to a clustering of multiple distinguishing features of a minority component. As a telling example, if in a ${\cal{D}}=30$-dimensional feature space one of the variables were identically repeated 10 times, and \texttt{RanBox} performed a search in ${\cal{D'}}=10$-dimensional subspaces, the algorithm would be very likely to end up focusing on the same narrow interval (any one would do) of each of those features: {\em e.g.}, a 10-dimensional box of width 0.1 in each of the correlated features would have a volume of $10^{-10}$; if there were $N=10,000$ events in the space, such a box would be predicted to contain $N_{exp}=10^{-6}$ events, while it would in fact contain exactly 1000 events!

\begin{figure}[h!]
\begin{center}
\includegraphics[width=12cm]{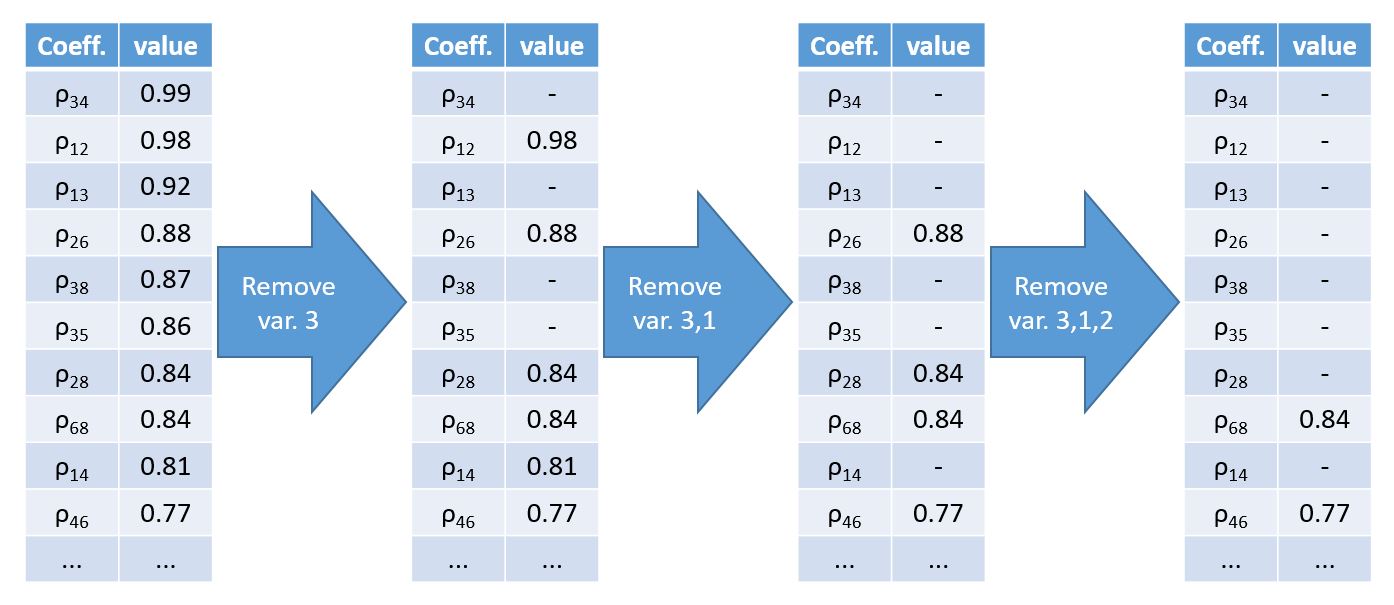}
\small
\caption{\small Graphical description of the CVR procedure available in the preprocessing stage of \texttt{RanBox} and \texttt{RanBoxIter}. The ordered list of absolute values of correlation coefficients among the variables defining the $\cal{D}-$dimensional feature space is scanned by searching for all possible combinations of $N_{void}$ variables which, once removed, minimize the largest surviving correlation coefficient. In the figure, for $N_{void}=3$ the removal of variables 3, 1, 2 (shown in succession for clarity) reduces the highest surviving correlation most effectively.}
\normalsize
\label{f:killcorr}
\end{center}
\end{figure}

Our correlated variable removal (CVR) procedure, which performs the identification of variables to be discarded, works as follows. We first compute the correlation coefficients $\rho_{ij}$ among all pairs of variables $ij$, and order them in a list by decreasing absolute value $|\rho_{ij}|$. Then we choose the number of variables to be removed $N_{void}$; in order to identify these we consider that if the $k$-th variable is removed, all correlation coefficients that include $k$ as one of the two indices will become irrelevant. We thus find the combination of $N_{void}$ variables which, when removed, minimizes the value of the highest surviving correlation coefficient. A graphical example of the technique is shown in Fig.~\ref{f:killcorr}.

%%%%%%%%%%%%%%%%%%%%%%%%%%%%%%%%%%%%%%%%%%%%%%%%%%%%%%%%%%%%%%
\subsection{Choices of a test statistic \label{s:teststatistic}}

\noindent
We consider two estimates of the expected number of events contained in a multi-dimensional region of the unit hypercube resulting from the standardization procedure, both corresponding to a binomial ratio. The first one is simply \par

\begin{equation}
    N_{exp,V} = N V_{box}.
    \label{eq:volexp}
\end{equation}

\noindent As the total copula space volume is $V=1$, the above estimate is only driven by the extension of the box volume $V_{box}$. The expectation results from assuming that the data distribute in the feature space with a constant density, and is useful in cases when $p_b(x)$ contains little structure in its copula, as departures from that assumption can then easily be associated with anomalous contaminations. This measure is the default one for the studies of algorithmic performance presented in Sec.~\ref{s:synthetic}, which are performed on synthetic datasets where the assumption above is identically true in the limit $f_s=0$.

\begin{figure}[h!]
\begin{center}
\includegraphics[width=10cm]{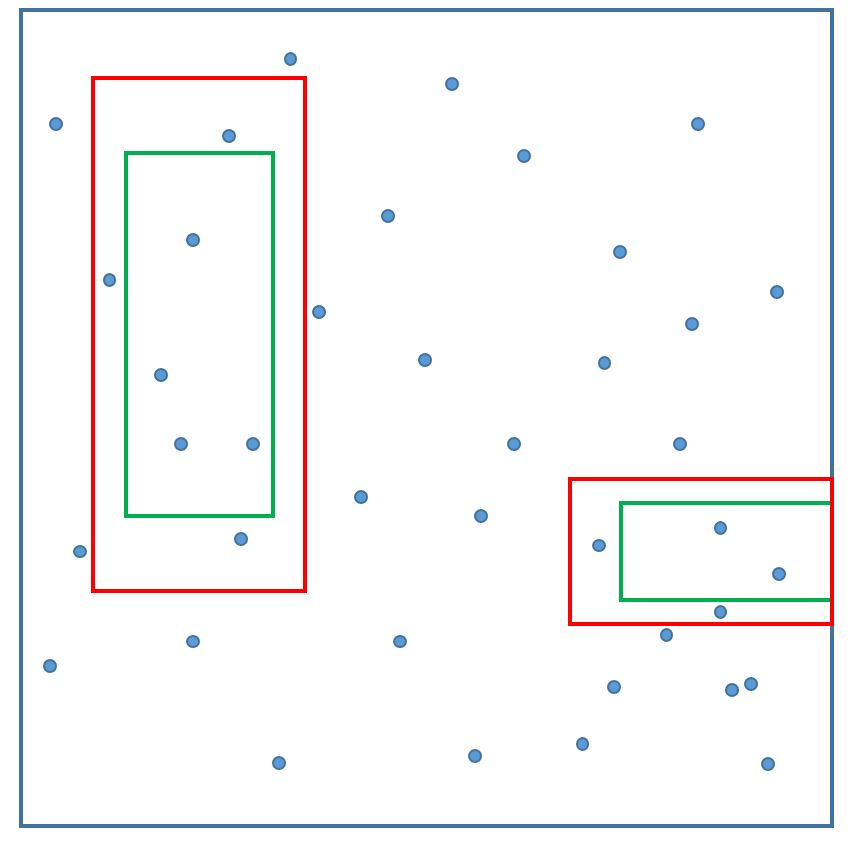}
\caption{\small Representation of the sideband method for box density estimation. Two possible search boxes in a 2-dimensional space are
shown in green; the relative sidebands, constructed according to the recipe of Eq.~\ref{eq:sideband_def}, are the regions between the red and the green rectangles. The sideband region on the lower right can only extend horizontally to the left, and the area it defines is thus  smaller than that of the related search box.}
\label{f:sidebands}
\end{center}
\end{figure}

\noindent
A second estimate, affected by higher statistical uncertainty than the former but conversely much less affected by a non-uniform density $p_{b}(x)$ in the copula space, may be obtained by defining a ``sidebands'' (SB) region that surrounds the search box (see Fig.~\ref{f:sidebands}). In this case, no reliance is made on overall constancy of the density for non-anomalous events, and the estimate leverages the density of data in the immediate neighborhood of the search box. If $[x^i_{min},x^i_{max}], i=1....\cal{D'}$ are the boundaries of the search box, the SB region is defined by the following relations:\par

\begin{equation}
    \delta_i = 0.5(x^i_{max}-x^i_{min})(2^{1/\cal{D'}}-1), 
\end{equation}
\begin{equation} 
     x^i_{min,SB} = \max(0,x^i_{min} - \delta_i),   
\end{equation}
\begin{equation} 
     x^i_{max,SB} = \min(1,x^i_{max} + \delta_i),
\label{eq:sideband_def}
\end{equation}

% GS Might be useful to show a figure of the volume + side-band in 2D --> added

\noindent
with $x^i \notin [x^i_{min},x^i_{max}]$ for at least one $i$, {\em i.e.} the SB volume does not include the search box volume. The SB then has a volume at most as big as the search box volume; it is in general smaller than that, as some of the intervals cannot extend on each side of the search box by the required length $\delta_i$, due to the hard boundaries at 0 and 1 (see again Fig.~\ref{f:sidebands}). If one observes a number of events $N_{out}$ in the sideband region, the expectation value of the number of events in the search box in the assumption of uniformity may be written as \par

\begin{equation}
    N_{exp,\tau} = \tau N_{out} ,
\end{equation}

\noindent
where 

\begin{equation}
    \tau = \frac{V_{box}}{V_{SB}}
\end{equation} 

\noindent is defined by the volumes of sideband region $V_{SB}$ and search box  $V_{box}$. 
A slight modification of the recipe for the expectation value above, which we have found to be effective, is operated when the number of observed sideband events $N_{out}$ is zero. In that case, which is frequent for large dimensionality searches and small statistics of the data sample, it is useful to reset $N_{exp}$ to the full-volume prediction, Eq.~\ref{eq:volexp}. We stick to this recipe in our applications of the sideband method in the studies described in this work.

To formulate the problem in its generality through the above definition of the extrapolation variable $\tau$, we observe that the full-volume estimate in Eq.~\ref{eq:volexp} corresponds to setting \par

\begin{equation} 
\tau = \frac{V_{box}}{1-V_{box}}  
\end{equation}

\noindent and $N_{out}=N - N_{in}$. In either case a likelihood-ratio-based test statistic may now be defined as follows: \par

\begin{equation}
    Z_{PL} = \sqrt{2} \bigg\{ N_{in} \ln \left[ (1+\tau) \left(\frac{N_{in}}{N_{in}+N_{out}} \right) \right] + N_{out} \ln \left[ \frac{1+\tau}{\tau} \left( \frac{N_{out}}{N_{in}+N_{out}} \right) \right] \bigg\}^{0.5}
\end{equation}

\noindent The above defined function has been shown~\cite{liandma} to be a good approximation of the Z-score corresponding to the binomial probability of observing an excess of events $N_{in}-N_{exp,\tau}$ in the box. It is to be noted, however, that $Z_{PL}$ cannot be considered a genuine signal significance, because in real applications ``non-anomalous'' data contain structure in the copula due to interdependence of their features; as a result, the $Z_{PL}$ test statistic for the null hypothesis has fatter tails at positive values than a Normal distribution. In addition, as discussed {\em infra} in more detail, \texttt{RanBox} effectively operates multiple testing on the dataset, hence $Z_{PL}$ cannot be used as a significance measure in the absence of a Bonferroni or similar correction~\cite{bonferroni}. Despite the above caveats, the fact that $Z_{PL}$ is a principled proxy to the significance of an excess in a Binomial counting experiment makes it a sound choice for a test statistic when the focus is the search for significant, anomalous signals.

We have observed that the $Z_{PL}$ test statistic is especially useful when anomalies are sought which may interest wide volumes of the feature space, with $N_{exp}$ correspondingly being not very small ---typically in the range of several tens to a hundred of events. Conversely, when the expectation $N_{exp}$ in the overdense region amounts to only a few events or less, an attractive alternative is to use the function $R_{reg}$ defined as \par
\begin{equation}
    R_{reg} = \frac{N_{in}}{N_{exp} +N_{reg}},
\end{equation}

\noindent with, {\em e.g.}, the regularization term set to $N_{reg}=1$. The maximization~\footnote{ In this work we stick to the setting $N_{reg}=1$, and consequently address the test statistic as $R_1$.} of $R_{reg}$ may identify more effectively small anomalies well confined in the search volume, in cases when the copula space of non-anomalous events has a rich structure, capable of producing high values of $Z_{PL}$ in regions of large volume and thus diverting the algorithm's attention from small, well-confined anomalies.

%%%%%%%%%%%%%%%%%%%%%%%%%%%%%%%%%%%%%%%%%%%
\subsection{Box seeding \label{clustering}}

\noindent
The search for the most overdense multi-dimensional interval in a feature space populated by sparse data points is complicated by the presence of a large number of local extrema; hence, a careful initialization of the box location and dimensions may significantly improve the performance of the algorithm. Although we tried several recipes for this task, here we only describe three of them, which we found the most suitable for our applications.

The baseline method, ``Algorithm 0'', consists in a fixed initialization of the box to a multi-dimensional interval of total volume $V_{box}$, set to equal a given fraction of the unit volume of the full feature space hypercube. The box, which lives in a $\cal{D}'$-dimensional subspace of the copula, is constructed by defining intervals $x^i_{min}$, $x^i_{max}$ (with $i=1,...,{\cal{D}}'$) as follows:\par

\begin{equation}
\Delta = \frac{1-V_{box}^{1/\cal{D'}}}{2}
\end{equation}
\begin{equation}
    x^i_{min} = \Delta 
\end{equation}
\begin{equation}
    x^i_{max} = 1 - \Delta  
\end{equation}

\noindent
An optimization of the initial value of $V_{box}$ is of course impossible in a unsupervised search, where neither non-anomalous or anomalous data have a specified density. However, our tests suggest that  setting $V_{box}=0.1$ is a reasonable choice when, as is the case in several of our considered applications, $\cal{D'}$ lays in the 6-10 dimensions range. {\em E.g.,} with ${\cal{D'}}=6$ one obtains starting intervals equal to $[0.16,0.84]$, and with ${\cal{D'}}=10$ intervals equal to $[0.10,0.90]$. Note that this corresponds to a relatively large box, in terms of its extension along each marginal. When combined with a  search algorithm that considers initial expansions or shrinkages in each of the box dimensions by amounts sufficient to extend all the way to the unit hypercube boundaries, the above initialization ensures that no overdensity laying close to the boundary of a coordinate will be overlooked by the search algorithm taking a step in the wrong direction at the start of the search.

\begin{figure}[h!]
\begin{center}
\includegraphics[width=12cm]{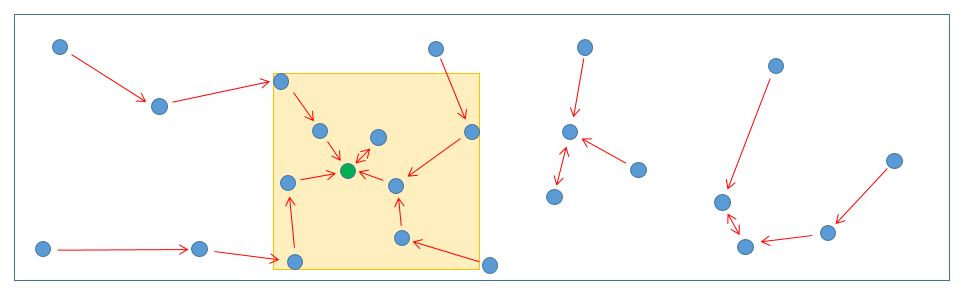}

\caption{\small Graphical description of the clustering algorithm used for box initialization with Algorithm 1. Blue points indicate the position of events in the two shown variables of the feature space. Arrows pointing from an event to another indicate the location of the closest neighbour of the event originating the arrow (according to a metric described in the text). The green point is the closest to four others, and it provides the seed of the box: the collection of all events which point to those four events define the boundaries of the box.}
\normalsize
\label{f:clustering}
\end{center}
\end{figure}

The second method, ``Algorithm 1'', is instead based on clustering the data based on a specialized Nearest-Neighbour (NN) search. First, the nearest neighbour $j$ is found for every event $i$ in the data, by using as a distance the following function: \par

\begin{equation}
    d_{ij} = \prod{_{k=1}^{{\cal{D'}}/2} |x_{o_k(ij)}^i-x_{o_k(ij)}^j|}
\end{equation}

\noindent where $o_k(ij)$ are the ${\cal{D'}}/2$ indices identifying the spatial coordinates for which the intervals $|x^i-x^j|$ are the smallest. In other words, the map $d_{ij}$ determines the minimum volume of a ${\cal{D'}}/2$-dimensional box that includes events $i$ and $j$. Once $d_{ij}$ is defined for all $i$ and $j$, one may compute for every event $i$ the number of neighbouring events $j=j_1 ... j_{N_{cl}}$ that have $i$ as their closest event according to that metric. The event $i_{maxNN}$ with the maximum number $N_{cl,maxNN}$ of such neighbours now allows to identify all $N_{2^{nd} \, order}$ events which have any of the $N_{cl,maxNN}$ events as their own nearest neighbours. The box can finally be initialized as the smallest ${\cal{D}}'$-dimensional interval that includes all the $N_{2^{nd} \, order}$ neighbours. A graphical description of the algorithm is provided in Fig.~\ref{f:clustering}.

\noindent
A third initialization method, ``Algorithm 2'', uses instead a kernel estimation of the density for the identification of starting box boundaries. The density is evaluated at the position of each of the $N$ events as the sum of $N$ $\cal{D}$-dimensional Gaussian distributions centered at the location of every event in the sample, and with equal diagonal covariance matrices $C=k^2 I_D$, with $I_D$ the $\cal{D}$-dimensional identity matrix and $k$ a tunable parameter which must be chosen according to the total dataset size and the dimensionality of the $\cal{D'}$ subspaces scanned by \texttt{RanBox}; its default value, used in the applications described in this work, is $k=0.2$. Once the point of highest density $x_{HD}$ is identified, the box is initialized as the multi-dimensional interval whose extension in each coordinate $x$ is \par

\begin{equation} 
[\max(x_{HD}- \delta_2,0.), \min(x_{HD}+ \delta_2,1.)], 
\end{equation}

\noindent
with the provision that if $x_{HD}$ is at less than $\delta_2$ distance from the boundary at 0 (1), the interval defaults to $[0.,2\delta_2]$ or $[1.-2\delta_2,1.]$, respectively. The default value of $\delta_2$ is 0.2; {\em e.g.} this corresponds, for a 10-dimensional subspace search, to initial boxes of volume equal to or smaller than 0.0001: the expected number of events within a 10,000-event sample contained in a random box of that volume is 1.0, which is a suitable starting point for the background expectation in the test statistic maximization. Given that the initialization provided by Algorithm 2 offers a good candidate for an overdense region, the focusing on a small initial region of feature space has been observed to be effective in the tested applications of our interest: those are in fact cases when a small, overdense region exists in the first place.

%%%%%%%%%%%%%%%%%%%%%%%%%%%%%%%%%%%%%%%%%%%%%%%%%%%%%%%%%%%%%%%%%%%%%%%%%
\subsection {Maximization of the test statistic \label{maximization}}

\noindent
A search for the multi-dimensional interval providing the highest value of the chosen test statistic (either $Z_{PL}$ or $R_{reg}$ as defined in Sec.~\ref{s:teststatistic} above) in a $\cal{D'}$-dimensional subspace of the feature space can be performed as follows.
\vspace{.2cm}

{\bf Step 1}: The initialization of the box is performed with the algorithm of choice. A set of step parameters are set to the starting value $\lambda_i=0.5 \, (i=1...{\cal{D'}})$. A loop counter $N_{GD}$ is set to zero.

{\bf Step 2}: Seven possible modifications are considered for each of the $\cal{D'}$ intervals defining the box: \par

\begin{center}
\begin{tabular}{ll}
$(x^i_{min})'$ & $(x^i_{max})'$ \\
\hline
max$(x^i_{min} - \lambda_i,0)$              & $x^i_{max}$ \\
min$(x^i_{min} + \lambda_i, x^i_{max}-\epsilon)$ & $x^i_{max}$ \\
$x^i_{min}$    & max$(x^i_{max} - \lambda_i, x^i_{min}+\epsilon)$ \\
$x^i_{min}$    & min$(x^i_{max} + \lambda_i,1)$ \\
max$(x^i_{min} - \lambda_i,0)$ &  max$(x^i_{max} - \lambda_i,\epsilon)$ \\
min$(x^i_{min} + \lambda_i,1-\epsilon)$  & min$(x^i_{max} + \lambda_i,1)$ \\
$r^i_{min}=$ min$(r_1,r_2)$ & $r^i_{max}=$ max$(r_1,r_2)$ \\
\end{tabular}
\end{center}

\noindent where $\epsilon$ is a parameter determining the coarseness of the algorithmic scan in the feature space, fixed in applications described in this work to $\epsilon=0.01$. In the last line the values $r_1$, $r_2$ determining a ``random jump'' in the $i$-th interval are random numbers sampled from a uniform distribution in $[0,1]$. The values of $(x^i_{min})'$ and $(x^i_{max})'$ defined above are rounded off to two decimal places in all cases. For each of these $7 {\cal{D'}}$ variations, an associated SB region is defined by the recipe described {\em supra}; this determines the numbers $N_{in}$ and $N_{out}$ and consequently the $7 {\cal{D'}}$ values of the test statistic of choice.

{\bf Step 3}: If the highest among the $7\cal{D'}$ values of the test statistic corresponding to the tentative box modifications is higher than the current maximum value, the box is modified to the corresponding new multi-dimensional interval, and all $\lambda_i$ values for the coordinates not affected by the change are reduced as follows: \par

\begin{equation}
    \lambda_i \to \max{(f \lambda_i,\epsilon)}
\label{eq:shrink}
\end{equation} 

\noindent
where the factor $f$ is set to $0.9$. In addition, if the box modification is chosen based on one of the $\cal{D'}$ random intervals $[r^i_{min},r^i_{max}]$, a counter $j_i$ is incremented by one; once a $j_i$ reaches a maximum value (10 by default), no more random jumps are allowed for the intervals in variable $i$. This recipe allows to control the convergence of the algorithm as well as the trade-off between its CPU consumption and its freedom in exploring new box configurations in the considered feature space dimensions.

If, instead, the current value of the test statistic is higher than all of the $7 {\cal{D'}}$ new values, no modifications to the box boundaries are applied, and $\lambda_i$ values are reduced as in Eq.~\ref{eq:shrink}. 

{\bf Step 4}: The loop counter $N_{GD}$ is incremented by one. If $N_{GD}$ reaches a limiting value (set to 100 by default) the algorithm stops; the algorithm also stops if all values $\lambda_i$ have reached the value $\epsilon$. Otherwise, steps 2, 3, and 4 above are repeated.

\vspace {.2cm}
\noindent
Despite its simplicity, the procedure described above typically converges in 30 to 50 iterations for ${\cal{D}}'=6-10$, which are typical values for the considered applications of fixed-subspace searches. 

%%%%%%%%%%%%%%%%%%%%%%%%%%%%%%%%%%%%%%%%%%%%%%%%%%%%%%%%%%%%%%%%%%
\subsection {Iterative scan of subspaces: \texttt{RanBoxIter}}

\noindent
A variation of the above algorithm, called \texttt{RanBoxIter}, is based on a different approach to the problem of scanning high-dimensional spaces. \texttt{RanBoxIter} performs a scan of the subspaces of the feature space incrementally, starting with two-dimensional subspaces and adding dimensionality gradually, until a maximum value ${\cal{D}}'_{max}$ is reached. The maximum dimensionality may be chosen such that it adapts to preconceptions on the behaviour of the anomalous signal sought, similarly to the choice of $\cal{D}'$ for \texttt{RanBox}.

\vspace{.2cm}
\noindent
\texttt{RanBoxIter} works as follows:

{\bf Step 1}: a loop is performed on all the ${\cal{D}}({\cal{D}}-1)$ combinations of pairs of features. For each combination, the corresponding two-dimensional subspace is considered, and a search for the box that maximizes the predefined test statistic of choice is performed, exactly as is done by \texttt{RanBox}. At the end of the ${\cal{D}}({\cal{D}}-1)$ scans, a list is compiled of those subspaces where the $N_{best}$ boxes yield the highest values of test statistic. A reasonable value for $N_{best}$ is in the 20-50 range.

{\bf Step 2}: For each of the $N_{best}$ subspaces, all possible choices for an additional feature are considered; the search for the box with the highest value of the test statistic is performed in each of the three-dimensional subspaces. At the end of the scans (which are at most $N_{best}({\cal{D}}-2)$, but typically fewer due to repetitions) a new list of the $N_{best}$ boxes is compiled. 

{\bf Step 3-${\cal{D}}'_{max}+1$}: A repetition of Step 2 is performed iteratively, each time incrementing by one the dimensionality of the considered subspaces, from 4 to ${\cal{D}}'_{max}$.

\vspace{.2cm}
\noindent At the end of the above procedure, the box of interest is the one which corresponds to the highest value of the test statistic among the $N_{best}$ ones returned by the algorithm; the full list may however be used for more detailed studies of the feature space.

Depending on the values of $\cal{D}$ and ${\cal{D}}'$/${\cal{D}}'$, as well as on the distinguishability of the anomalous signal, \texttt{RanBoxIter} may take a shorter or a longer time to run than \texttt{RanBox} to reach comparable sensitivity: when the anomaly involves many distinguishing features the latter may require few scans of subspaces to find anomalous regions of feature space, while the former must consider a significant number of combinations of up to ${\cal{D}'}_{max}$ dimensions. For a comparison we consider a typical situation of interest for our applications, involving a 20-dimensional feature space and an anomalous overdensity involving a significant modification of $p_b(x)$ for 10 of the features. Let us further assume that the \texttt{RanBox} search is run on subspaces of 8 dimensions. In that case, the probability that a random choice of eight dimensions captures a significant number of features in which the signal is distinguishable will be small. The formula that yields the probability that a choice of C features out of D selects X of them within a subset S (with S $\geq$ X) is \par

\begin{equation}
    P(X,C,S,D) = \frac{c(S,X)c(D-S,C-X)}{c(D,C)}=
\end{equation}
\begin{equation}
    =    \frac{S!C!(D-C)!(D-S)!}{X!(S-X)!D!(C-X)!(D+X-S-C)!} 
\end{equation}

\noindent 
where $c(n,k)=\frac{n!}{k!(n-k)!}$ is the binomial coefficient for $(n,k)$.
With D=20, S=10, C=8, X=8 the odds are 3/8398, {\em i.e.} one in 2800; if X=7, this raises to 40/4199, {\em i.e.} about one in 105; for X=6, it reaches 315/4199, or about one in 13.3. Given these numbers, a \texttt{RanBox} run on a random choice of 1000 8-dimensional subspaces will encounter a large number of situations where the signal is distinguishable. Conversely, a run of \texttt{RanBoxIter} that sequentially scans subspaces of dimensionality from 2 to 8, each time considering the $N_{best}=20$ subspaces returned by the previous step, will have to scan a much larger number of combinations; although that number depends on the specific combinations of best boxes at each iteration, in practical situations it is of the order of 10,000. On the other hand, low-dimensional subspace scans will take a shorter time to be completed. 

In order to appraise the merits and peculiarities of the two algorithms it is not trivial, nor necessary, to go beyond the simple comparison offered {\em supra} of the number of combinations of subspaces that the algorithms consider. A much better option is to empirically observe their performance in controlled cases; this is what we present in the next Section.

%%%%%%%%%%%%%%%%%%%%%%%%%%%%%%%%%%%%%%%%%%%%%%%%%%%%%%%%%%%%%
\section{Performance studies with synthetic data \label{s:synthetic}}

%%%%%%%%%%%%%%%%%%%%%%%%%%%%%%
\subsection {Event generation} 

\noindent 
A synthetic dataset sampled from a multi-dimensional Uniform distribution $p_b(x) = {\cal{U}}(x)$, with $x \in [0,1]^{\cal{D}}$, may be generated by repeated calls to the TRandom3$\to$Uniform() routine~\footnote{ The random generation is based on the Mersenne primes, and has a periodicity of about $10^{6000}$.} of the ROOT package~\cite{root}, which we employ in our \texttt{c++} implementations of \texttt{RanBox} and \texttt{RanBoxIter}. Such a dataset may be considered the ideal background for an anomaly search: by lacking any internal structure in the copula, it constitutes a best-case scenario for performance evaluations of the algorithm in a controlled setting. 
%Having set the boundaries of each feature in $[0,1]$, the data fits in a unit hypercube, which is convenient for density estimations based on volume of subspaces. 
The unknown signal may instead be generated by drawing samples from a multi-dimensional Gaussian distribution in a subset ${x_g, g=1,...,N_g}$ of the features, $x_g \in {\mathbf{R}}^{N_g}$, and the remaining ones $x_u, u=N_g+1,...,\cal{D}$ from a uniform density. While Gaussians have support on the real axis, the generation ensures that the drawn features are also contained in the $[0,1]$ interval, as detailed below.

%\vspace{.2cm}
\clearpage
\noindent
We define the following default set of parameters:\par

\begin{itemize}
    \item (for background) $x_i = {\cal{U}}(0,1)$;
    \item (for signal) $x_u = {\cal{U}}(0,1)$;
    \item sigma $\sigma_{gg} = {\cal{U}}(0.01, 0.1)$;
    \item mean  $\mu_g = {\cal{U}}(3 \sigma_{gg}, 1 -3\sigma_{gg})$;
    \item $r_{gh} = {\cal{U}}(-1,1)$ (with $g,h \in \{1,...,N_g\}, g \neq h$).
\end{itemize}

\noindent
A random choice of $\sigma_{gg}$ and $r_{gh}$ values as defined above will not in general generate a positive-definite covariance matrix $C$ with variances $\sigma_{gg}^2$ and $\sigma_{gh}^2=r_{gh} \sigma_{gg} \sigma_{hh}$; hence the procedure of generating $C$ is repeated until a Cholesky-Banachiewicz (CB) decomposition $L L^T = C$ into a lower-triangular matrix $L$~\cite{cholesky} is found, which guarantees the positive-definite nature of $C$.
%~\footnote {At every iteration, we reduce their sampled range of correlation coefficients $r_{gh}$ by $0.1\%$, gradually increasing the chance to obtain a positive-defined matrix; {\em e.g.,} for 20-dimensional matrices the procedure converges after an average of $1000$ trials, with maximum correlation coefficients usually in the $[-0.3,0.3]$ range.}. 
Once successful, the CB decomposition allows to easily draw samples from the multi-dimensional Gaussian distribution by posing, for every $g$, \par

\begin{itemize}
    \item (for signal) $x_g = \mu_g + \sum_{h=1..N_g} L_{gh} n_h$ 
\end{itemize}

\noindent
with $n_h$ sampled from a Normal distribution. During event generation, if a coordinate sampled from the multivariate Gaussian exceeds the range $[0,1]$, it is simply resampled. This truncation has the effect that Gaussians with $\mu_g$ values close to the boundaries have an up to twice higher local density than Gaussians closer the center of the $[0,1]$ interval. For this reason, in most tests we limit $\mu_g$ values to the range stated above, except when we explicitly study the performance at the edges (see {\em infra}). 
%mirrored around the corresponding Gaussian mean, by posing $x_{g,[0,1]} = 2 \mu_g - x_g$. The resulting marginals, for $\mu_g$ values (1-$\mu_g$ values) comparable to the respective $\sigma_g$, take the shape of a Gaussian distribution with a truncation at 0 (1, respectively) and a corresponding "fat tail" with twice the density of a regular Gaussian tail for $x_g>2\mu_g$ ($ x_g<2\mu_g-1$, respectively). This truncation of the tails ensures that all data are contained in a unit hypercube, and preserve the correspondence between the width of box intervals and the corresponding integral of the signal component captured by it.
%For studies of a less distinguishable signal component (referenced as ``wide gaussians'' {\em infra}), characterized by a larger footprint in the feature space, we widen the sampling interval of the Gaussian sigma parameters to $\sigma_{gg} = {\cal{U}}(0.05,0.15)$. 
Although the background is already generated with flat marginals, after the inclusion of signal we of course re-standardize the dataset by using Eq.~\ref{eq:standardize1}.

\begin{figure}[h!]
\begin{center}
\includegraphics[width=12.5cm]{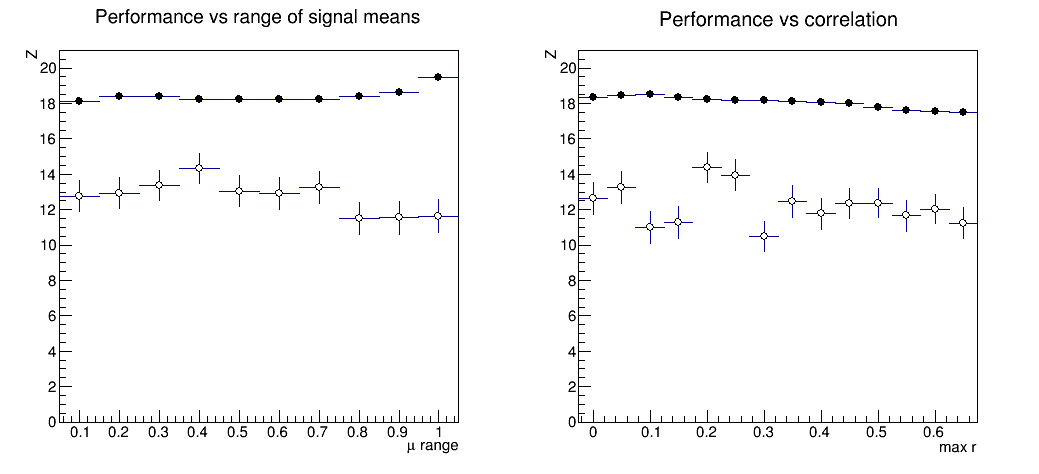}

\caption{\small Mean values of the test statistic $Z_{PL} $ as a function of the characteristics ($\Delta \mu_g$ for $r_{gh}=0.2$ (left), and $r_{gh}$ for $\Delta \mu_g=0.7$ (right)) of the signal component, from 50 repetitions of searches in synthetic datasets each composed of 50 signal events and 4950 background events. Black points correspond to searches initialized with Algorithm 2, empty points correspond to searches initialized with Algorithm 0. For reference, the critical region (for $\alpha=0.05$) corresponds to $Z_{PL}=7.1 (7.2)$ for Algorithm 0 (2, 
respectively). See the text for other detail.}
\normalsize 
\label{f:varymeansandrhos}
\end{center}
\end{figure}

When performing power tests of the algorithm, we avoid the random effect of varying $\sigma$ parameters, and use reference samples with a more narrowly defined signal component, by fixing all Gaussian sigmas to $\sigma_{gg}=0.05$. In this case correlation coefficients $r_{gh}$ are chosen at random within the discrete set $\{-max(r_{gh}), 0., max(r_{gh})\}$ by posing $max(r_{gh})=0.2$, and we allow means $\mu_g$ to vary at random in their default range, $[0.15,0.85]$. The different signals that correspond to varied means and correlations have equal chance of being identified by the algorithm. For example, Fig.~\ref{f:varymeansandrhos} shows average $Z_{PL}$ values from runs of the algorithm with the following choice of parameters:\par

\begin{itemize}
    \item $N_{b} = 4950$ background events
    \item $N_{s} = 50$ signal events
    \item ${\cal{D}} = 20$ active dimensions of feature space
    \item $N_G = 6$ Gaussian features in signal component
    \item ${\cal{D}}'=6$ dimensions for box definition
    \item $N_{trials} = 1$ subspace sampled per dataset, of features coincident with the $N_G$ in which signal component has a Gaussian distribution. 
    \item $N_{rep} = 50$ datasets generated and searched
    \item Algorithm 0 (random box initialization) and 2 (kernel density) used
    \item No dimensionality reduction (PCA or correlated features removal) performed 
\end{itemize}

\noindent 
By only considering, through the above choices, the subspace which yields the highest probability of locating a signal-rich box, we reduce the effect of randomness and allow for a more precise study of the impact of the tested parameters. In Fig.~\ref{f:varymeansandrhos} the values of the test statistic $Z_{PL}$ appear stable as a function of the sampled ranges $max(r_{gh})$ and $\Delta \mu_g = \mu_g^{max}-\mu_g^{min}$, indicating that the search algorithm is capable of locating overdensities regardless of their position in the space~\footnote{ The observed slightly lower performance of searches initialized by Algorithm 0 for $\Delta \mu_g$ values close to 1 is an effect of the higher chance of central signals to be initially contained in randomly-initialized boxes. Instead, Algorithm 2 allows to exploit the slightly higher maximum density reached by signals with one or more features close to the boundaries of the space, due to the already mentioned truncation we operate outside the $[0,1]$ range.}, and that the correlation between Gaussian-distributed variables does not affect the chance of identifying overdense multi-dimensional intervals. Similar results are obtained by initializing the box dimension with Algorithm 1 (kNN-seeded clustering), and/or by using $R_1$ as a test statistic.

%%%%%%%%%%%%%%%%%%%%%%%%%%%
\subsection{Power tests of \texttt{RanBox}} 

\noindent 
While in a unsupervised search one cannot in general define a hypothesis test, given the absence of hypotheses for the sampling distributions, we are still interested in verifying the ability of \texttt{RanBox} to locate overdense regions of the feature space as a function of its free parameters for a set of different benchmark datasets. This will enable a comparison of the fixed subspace search strategy to the iterative one, as well as provide a scale of the algorithm sensitivity. Hence we construct a ``flat'' dataset containing events uniformly distributed in the feature space, and ``signal'' datasets where a fraction of the events are sampled from a PDF which includes, for some of the features, a multivariate Gaussian component (see {\em supra}). Once a type-I error rate $\alpha$ is defined, the tail integral of the test statistic distribution $f(TS|H_1)$, output by \texttt{RanBox} searches on alternative hypotheses $H_1$ corresponding to datasets contaminated with events having multivariate Gaussian features, allows to construct a power function $1- \beta(\alpha)$ as \par
\begin{equation}
    1 - \beta(\alpha) = \int_{x_{cr}(\alpha)}^{\infty} f(x|H_1) dx ,
\end{equation}

\noindent
where $x_{cr}(\alpha)$ is defined by the relation \par

\begin{equation}
    \alpha = \int_{x_{cr}(\alpha)}^{\infty} f(x|H_0) dx .
\end{equation}

\noindent
To check the performance of the algorithm in a controlled setting, we define signal parameters by fixing the Gaussian sigma values in signal events to $\sigma_{gg}=0.05$, and allow means and correlations to vary in the range $\mu_g \in [0.15,0.85]$ and $r_{gh} \in$ \{-0.2,0.,0.2\}, respectively. We consider again samples of 5000 events, and study the power $1-\beta$ for the three choices $\alpha=0.05,0.01,0.001$, using ${\cal{D}}=20$ space dimensions. We also set the following algorithm hyperparameters:\par
\begin{itemize}
    \item Algorithm = 0
    \item $N_{trials}=1000$ subspaces scanned for each dataset
    \item test statistic used: $Z_{PL}$
    \item expectation value of events in the box: $N_{exp,V}$.
\end{itemize}

\begin{figure}[h!]
\begin{center}
\includegraphics[width=8cm]{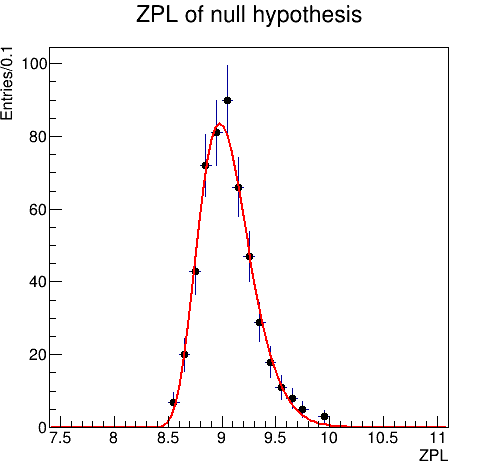}
\small
\caption{\small Distribution of the $Z_{PL}$ test statistic for 500 repetitions of \texttt{RanBox} tests of the null hypothesis in 5000-event background-only samples; a fit to a Gamma function is overlaid. 1000 subspaces are scanned with Algorithm 0 for the box initialization. See the text for other details.}
\normalsize
\label{f:TS_H0}
\end{center}
\end{figure}

\begin{figure}[h!]
\begin{center}
\includegraphics[width=12.5cm]{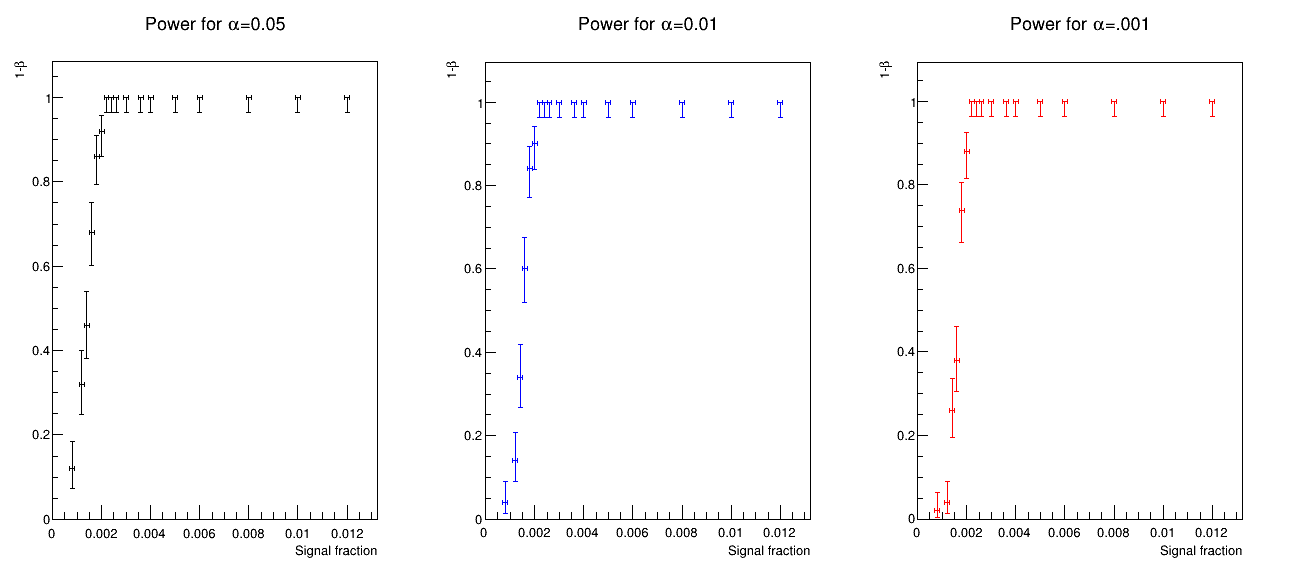}
\includegraphics[width=12.5cm]{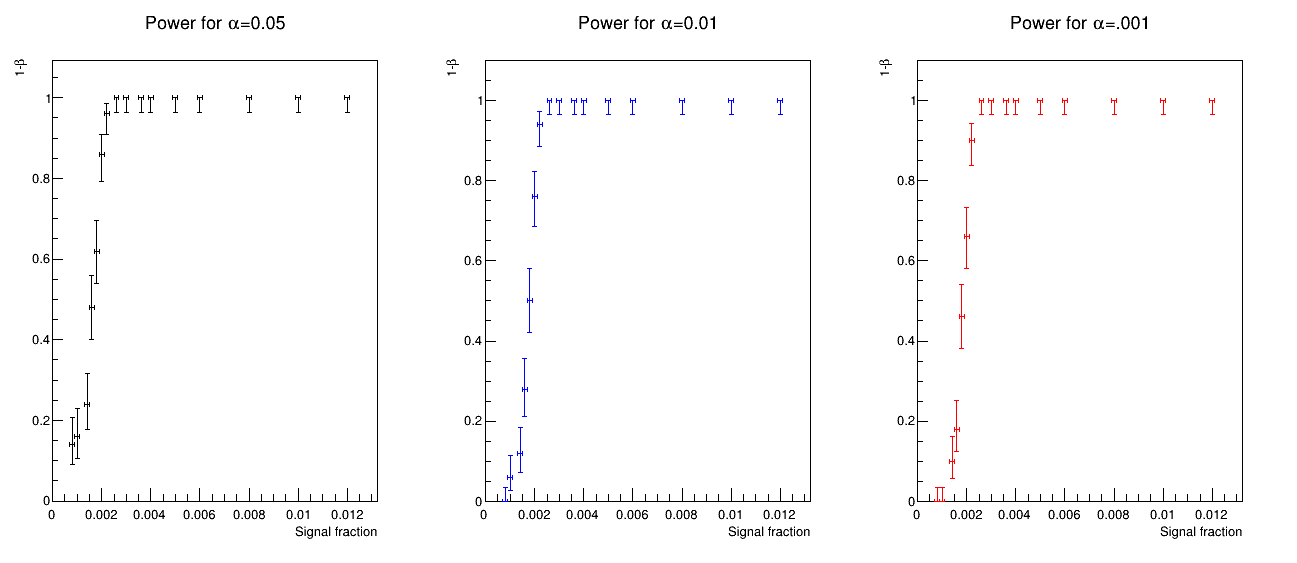}
\caption{\small  \texttt{RanBox} (top) and \texttt{RanBoxIter} (bottom) power curves for $Z_{PL}$ as a function of the fraction of signal in 5000-event samples. The black points (left) correspond to $\alpha=0.05$, the blue points (center) to $\alpha=0.01$, and the red points (right) to $\alpha=0.001$; the critical region for the latter two tests are obtained from extrapolated values of $Z_{PL}$ for the null hypothesis. $68.3\%$ intervals are computed with the Clopper-Pearson method for the Binomial ratio. 
See the text for other details.}
%1000 subspaces are scanned, where box dimensions are initialized with Algorithm 0; see the text for other details.}
\label{f:power_varySF}
\label{f:power_varySF_iter}
\end{center}
\end{figure}

\noindent
In a first test we fix the number of features where the signal component exhibits a Gaussian distribution to $N_g=15$, and vary the number of signal events in the generated samples. The critical region is directly obtained for $\alpha=0.05$ from the distribution $f(TS|H_0)$ obtained by repeating 500 times the procedure of generation and 1000-subspace-search of datasets including no signal. For the two smaller values of $\alpha$ ($0.01$, $0.001$), we instead rely on the modeling of the distribution of $f(TS|H_0)$ with a Gamma function (see Fig.~\ref{f:TS_H0}) to determine the corresponding $x_{cr}$ values. For each studied value of the signal component we obtain 50 values of $f(TS|H_1)$, from which we extract the power as the fraction of values in the critical regions corresponding to the three chosen values of $\alpha$. The results of this test are shown in Fig.~\ref{f:power_varySF} (top row). We observe that \texttt{RanBox} is fully capable of spotting localized accumulations due to a multivariate Gaussian signal, down to few-per-mille contaminations of the data sample.

\begin{figure}[h!]
\begin{center}
\includegraphics[width=12.5cm]{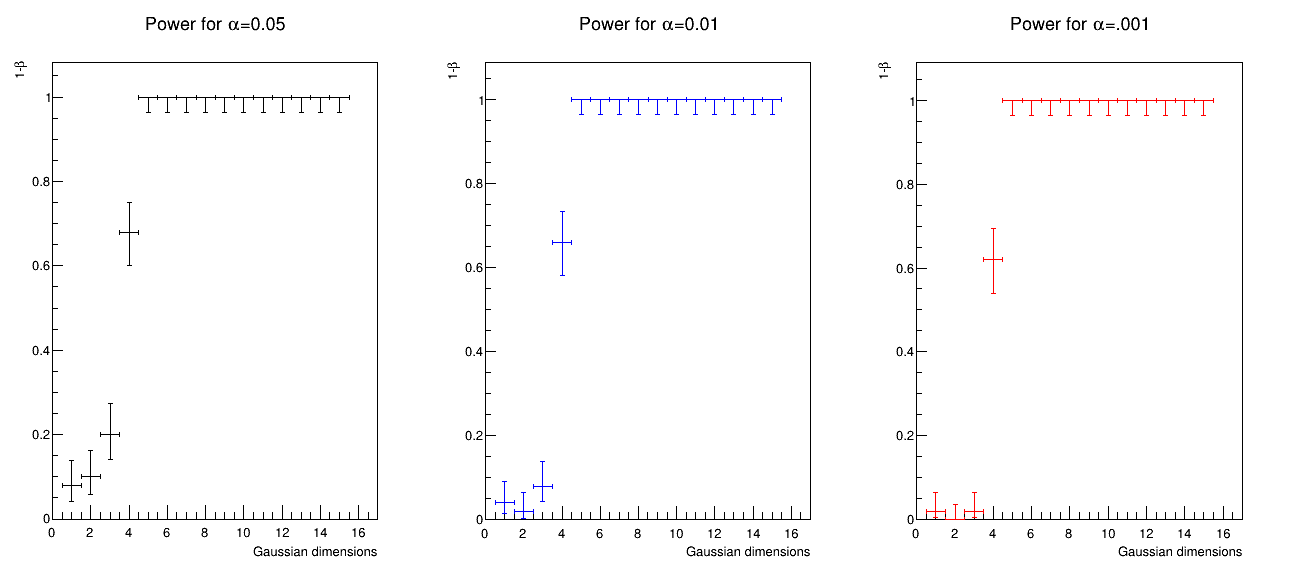}
\includegraphics[width=12.5cm]{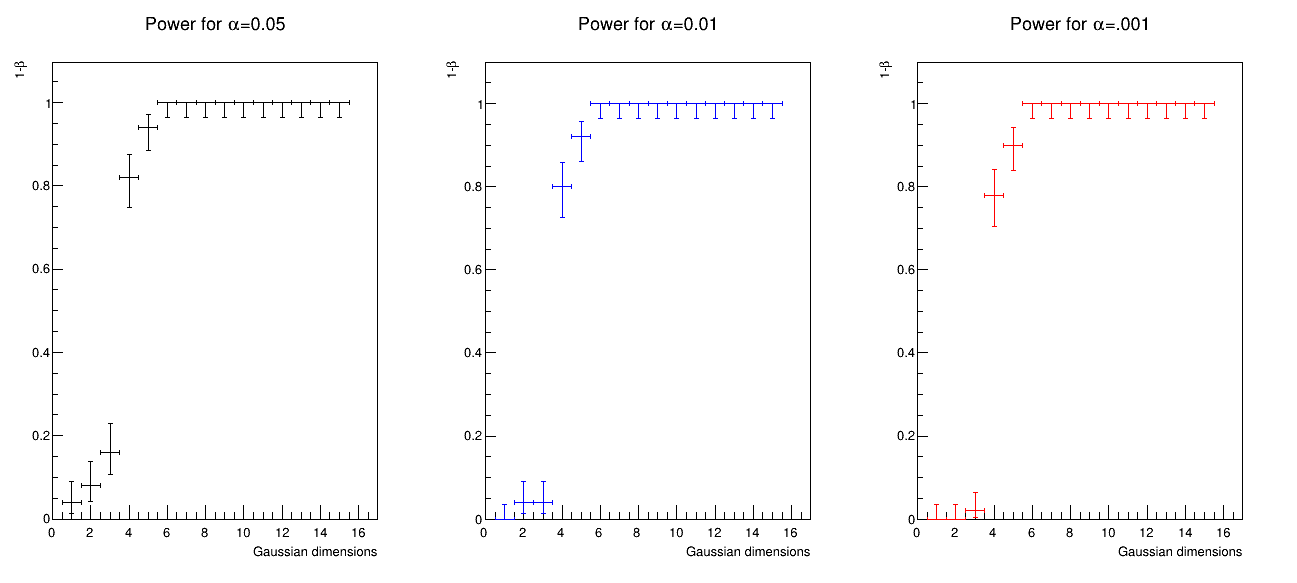}
\small
\caption{\small \texttt{RanBox} (top) and \texttt{RanBoxIter} (bottom) power curves for $Z_{PL}$ as a function of the number of Gaussian features in signal events, in samples containing 50 signal events and 4950 flat-distributed events. The black points correspond to $\alpha=0.05$, the green points to $\alpha=0.01$, and the red points to $\alpha=0.001$; the latter two are obtained from extrapolated values of the critical region. $68.3\%$ intervals are computed with the Clopper-Pearson method for the Binomial ratio. See the text for other details.} 
%1000 subspaces are scanned, where box dimensions are initialized with Algorithm 0; see the text for other details.}
\normalsize
\label{f:power_varyng}
\label{f:power_varyng_iter}
\end{center}
\end{figure}

\begin{figure}[h!]
\begin{center}
\includegraphics[width=12.5cm]{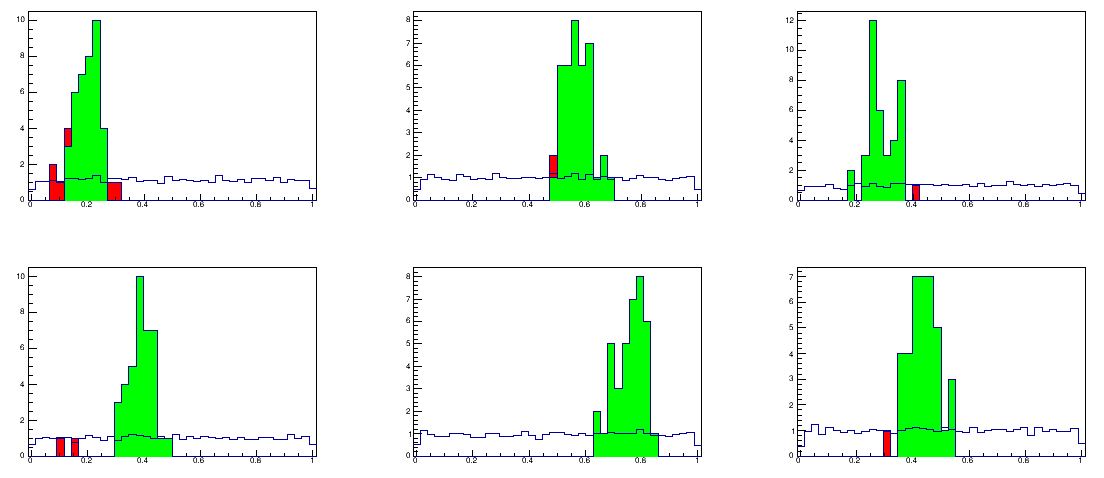}
\caption{\small Distribution of the six features defining the subspace where \texttt{RanBox} finds the highest-$Z_{PL}$ box in a run on 5000 synthetic events, 4950 of them generated from a $D=20$-dimensional uniform distribution and the remaining 50 ``signal'' events generated with 11 features drawn from a multidimensional Gaussian distribution. The blue histograms show the totality of the data; the filled green histograms show the distribution of events contained in the highest-$Z_{PL}$ box; the filled red histograms show the distribution of events that fail to be contained in the box only because of their value on the displayed variable. See the text for other details. }
\label{f:histo_11g}
\end{center}
\end{figure}

In a second study we determine, with the same procedure described {\em supra}, the power of \texttt{RanBox} as a function of the number of Gaussian dimensions $N_g$ of the signal component, by fixing the signal fraction to $f_s=1\%$ ({\em i.e.}, 50 signal events and 4950 background events). We observe in Fig.~\ref{f:power_varyng} (top row) that there is sensitivity to multivariate Gaussian signals that involve even only few (4 and above) of the 20 dimensions of the feature space.

\begin{figure}[h!]
\begin{center}
\includegraphics[width=12.5cm]{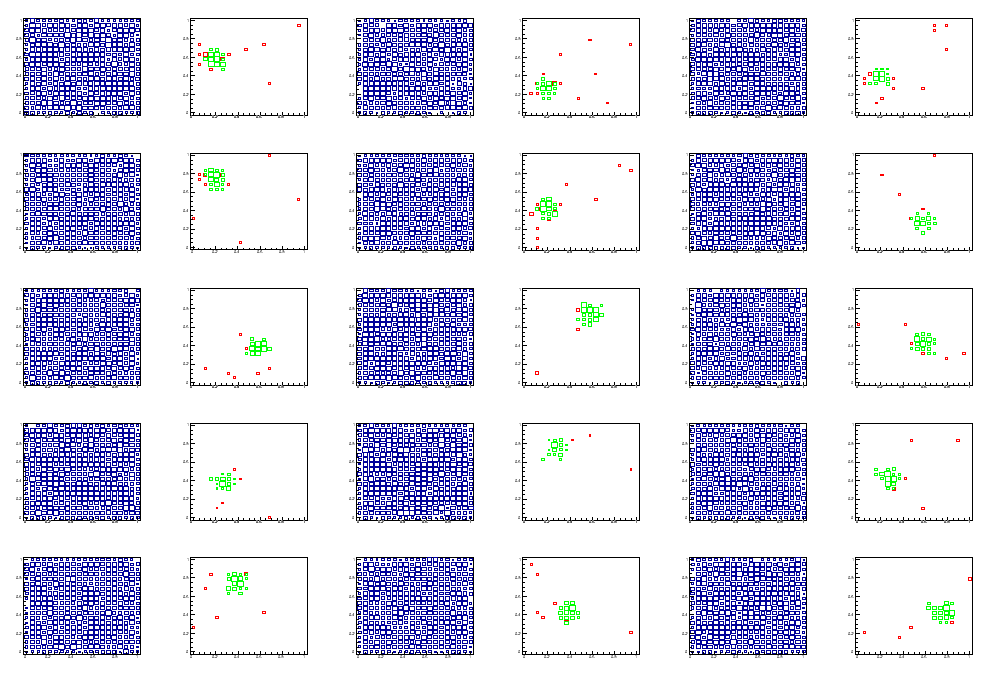}
\small
\caption{\small Scatterplots of the six features defining the subspace where \texttt{RanBox} finds the highest-$Z_{PL}$ box in a run on 5000 synthetic events, 4950 of them generated from a $D=20$-dimensional uniform distribution and the remaining 50 ``signal'' events generated with 11 features drawn from a multidimensional Gaussian distribution. The distribution of the totality of the data is shown in blue on the left of each pair of graphs, while the distribution of selected events (in green) is shown in green on the corresponding right graph; in red are events that fail to be included in the highest-$Z_{PL}$ box only because of their value of the shown features. From top to bottom and left to right each pair of graph describes the spaces $(v_1,v_2)$, $(v_1,v_3)$, $(v_1,v_4)$ (first row), $(v_1,v_5)$, $(v_1,v_6)$, $(v_2,v_3)$ (second row), $(v_2,v_4)$, $(v_2,v_5)$, $(v_2,v_6)$ (third row), $(v_3,v_4)$, $(v_3,v_5)$, $(v_3,v_6)$ (fourth row), and $(v_4,v_5)$, $(v_4,v_6)$, $(v_5,v_6)$ (fifth row). See the text for other details. }
\normalsize
\label{f:scatter_11g}
\end{center}
\end{figure}

In Fig.~\ref{f:histo_11g} and Fig.~\ref{f:scatter_11g} we provide a visualization of sample results of a \texttt{RanBox} run. The first figure shows marginal distributions of the six features where \texttt{RanBox} identifies an anomalous signal, in the copula space (where the total dataset has by definition uniform marginals before the selection). The subspace where the best box is found is one where the signal exhibits Gaussian distributions in all the features, and all the events in the box are in fact due to the signal component. The scatterplots of Fig.~\ref{f:scatter_11g} show two-dimensional distributions of the full data sample and the data selected as the best box. This further demonstrates the correct working of the algorithm, which can effectively extract the overdense region from an apparently flat distribution. The conclusions we draw are that the algorithm performs as expected when run on a synthetic data sample and in controlled conditions.

%%%%%%%%%%%%%%%%%%%%%%%%%%%%%%%%%%%%%%%%%%%%%%%%%
\subsection{Power tests of \texttt{RanBoxIter}}

\noindent
We follow the same approach as the one discussed {\em supra} for tests of the power of \texttt{RanBoxIter} on synthetic datasets. In Fig.~\ref{f:power_varySF_iter} (bottom row) we show the power of hypothesis tests based on the most significant box of dimension up to 12 found by the iterative algorithm, as a function of the signal fraction in 5000-event samples and for three different type-I error rates. In those tests, the signal has 15 Normally-distributed features out of 20, with the same settings for width, location, and correlation among them already described above. In Fig.~\ref{f:power_varyng_iter} (bottom row) we then show the result of varying the number of Gaussian dimensions of the signal, in datasets with $1\%$ signal fraction (50 signal events and 4950 background events), again for three different type-I error rates.

From inspection of the two sets of graphs we verify that the iterative algorithm is effective in identifying the signal in these controlled conditions. We also observe that its power is comparable to that of the non-iterative version. Of course, as the performance of \texttt{RanBox} strongly depends on the number of scanned subspaces, and conversely that of \texttt{RanBoxIter} on the value of $N_{best}$, one may not draw general conclusions, and computing load considerations have to be included in a choice of which version of the algorithm to employ for generic applications. In addition, since the performance of \texttt{RanBoxIter} is reliant on the existence of two-dimensional subspaces where the signal component produces overdensities, we believe that rather than making an {\em a priori} choice of which version to employ, users should consider scanning their datasets with both methods for a deeper insight.

\clearpage
%%%%%%%%%%%%%%%%%%%%%%%%%%%%%%%%%%%%%%%%%%%%%
\section {Experiments \label{s:experiments}}

\noindent
The power tests described in Sec.~\ref{s:synthetic} are about as far as one can go to characterize the performance of \texttt{RanBox} and \texttt{RanBoxIter}, since on any real-life dataset the specificities of the data structure and the lack of generalization power of the algorithm will make it pointless to investigate in a systematic way its optimal settings and resulting sensitivity. For this reason, in this Section we free ourselves of the need to assess confidence intervals on all the reported statistics, which would also entail a quite significant computing burden~\footnote{ The tests we report in this work overall cost several thousand hours of single-machine CPU by themselves. }, and prefer to offer sets of results of single runs of the algorithm on samples of data taken from three datasets, the first two offered by particle physics research and the third taken from an industrial application. This will allow us to walk the reader through the possible uses and search methodology that can be adopted in addressing the search for anomalies.

%%%%%%%%%%%%%%%%%%%%%%%%%%%%%%%%%%%%%%%
\subsection{Exotic signals in LHC data}

\noindent
The search of new phenomena in LHC proton-proton collisions data is the very application that \texttt{RanBox} is designed to address. A signal of new physics may manifest itself as a localized increase in density in some of the features derived from particle interactions in the detector. A model-independent search should consider a complete set of kinematical features describing the observed particles in the final state of the collision events, and perform an unbiased scan of their combined multi-dimensional distribution.

For a test of \texttt{RanBox} on the above use case we rely on the large dataset of simulated proton-proton collisions available in the University of Irvine's repository~\cite{uci}, a dataset known by its nickname ``HEPMASS''. This dataset was generated explicitly to test multivariate algorithms for classification and search of small signals in large background datasets. The generated signal is that of an exotic resonant particle X, with a mass of 1000 GeV, which decays to a pair of top quarks, $X \to t \bar{t}$, when the top quarks successively produce in their decay a single-lepton final state characterized by a high-energy electron or muon, a neutrino, and four hadronic jets. Background samples describe all Standard Model processes that produce a similar final-state signature. The ATLAS experiment is considered as the detector that performs the reconstruction of the produced particle signals; more detail on the generated dataset and the simulation are available in~\cite{whiteson}.

\begin{table}[h!]
    \centering
    \begin{tabular}{r|l|l}
    Number &  Feature & Description \\
    \hline
    1-3    &  $P^{\ell}_{xyz}$   & Primary lepton momentum \\
    4-6    &  $P^{j_1}_{xyz}$  & First jet momentum \\
    7      &  $P^{j_1}_{tag}$  & First jet b-tag information \\
    8-10   &  $P^{j_2}_{xyz}$  & Second jet momentum \\
    11     &  $P^{j_2}_{tag}$  & Second jet b-tag information \\
    12-14  &  $P^{j_3}_{xyz}$  & Third jet momentum \\
    15     &  $P^{j_3}_{tag}$  & Third jet b-tag information \\
    16-18  &  $P^{j_4}_{xyz}$  & Fourth jet momentum \\
    19     &  $P^{j_4}_{tag}$  & Fourth jet b-tag information \\
    20     &  $P_T^{miss}$& Missing transverse momentum \\
    21     &  $\phi_{P_T^{miss}}$ & Missing transverse momentum azimuthal angle \\
    \hline
    22     &  $m_{\ell \nu}$ & Mass of reconstructed lepton-neutrino system \\
    23     &  $m_{jj}$    & Mass of jets from $W \to q q'$ decay products \\
    24     &  $m_{jjj}$   & Mass of reconstructed $t \to Wb \to bqq'$ decay system \\
    25     &  $m_{j \ell \nu}$ & Mass of reconstructed $t \to Wb \to l \nu b$ decay system \\
    26     &  $m_{Wbb}$    & Mass of leptonic W plus leading b-jets \\
    27     &  $m_{WWbb}$   & Mass of hypothetical $X \tt$ resonance \\
    \hline
    \end{tabular}
\small
    \caption{\small List of the 27 features of signal and background events in the HEPMASS dataset. The first 21 are low-level features, the last 6 are higher-level ones produced by combining the low-level features into physics-motivated observables. See the text for more detail.}
    \normalsize
    \label{t:featuresHEPMASS}
\end{table}

\noindent
The data are characterized by reconstruction-level variables from a fast simulation. An idealized reconstruction of a proton-proton collisions yielding top quark pairs is performed, identifying the observed jets, leptons, and b-jets~\footnote{We call ``b-jet'' a hadronic jet which has been originated from a b-quark. When classified as such by a software algorithm, the jet is said to be ``b-tagged''. }. From the reconstruction of the event, the low-level kinematic features obtained are particle momenta: the momentum of the leading lepton, the momentum of the four leading jets (in decreasing order of transverse momentum) and related b-tagging information, and magnitude and azimuthal angle of the so-called ``missing transverse momentum'' vector. The latter is defined as the opposite of the sum of the momentum vectors of all observed particles, calculated in the transverse plane of the particle beams~\footnote{ Missing transverse momentum carries information on the momenta of neutrinos, particles typically produced in weak boson decays that do not leave a traceable signal in the detector but can still be inferred from the imbalance of the momenta of observed particles.}.

\begin{figure}[h!]
    \begin{center}
    \includegraphics[width=12cm]{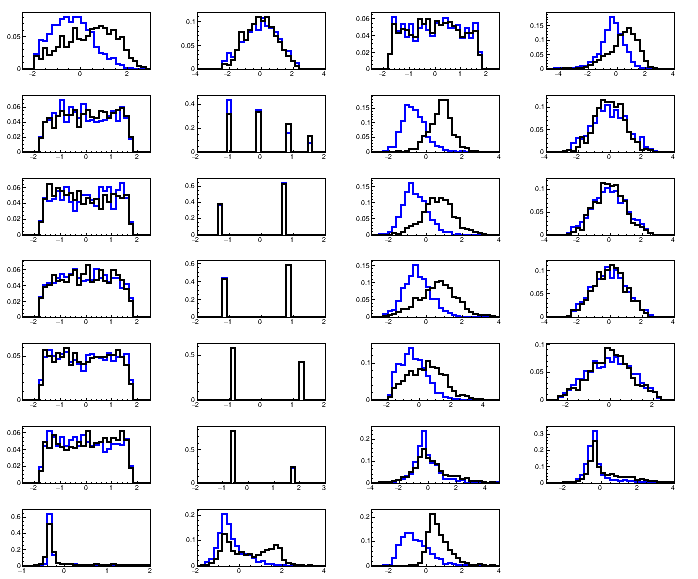}
\small
    \caption{\small Normalized and standardized distributions of the 27 features of HEPMASS data for signal (black) and background (blue).}
    \normalsize
    \label{f:featuresHEPMASS}
    \end{center}
\end{figure}

\noindent
The high-level features of the set are the values of the invariant masses of the intermediate objects calculated using the low-level kinematic features, in the hypothesis that a correct identification of decay objects and assignment to final state particles has been obtained. These are: $m_{\ell \nu}$ from the decay process $W \rightarrow \ell \nu$, $m_{j j}$ from the $W \rightarrow q q^{\prime}$ process, $m_{j j j}$ from the $t \rightarrow W b \rightarrow b q q^{\prime}$ process, $m_{j \ell \nu}$ from the $t \rightarrow W b \rightarrow \ell \nu b$ process, and the combined $m_{W W b b}$ mass of the decay products assumed for X. Table~\ref{t:featuresHEPMASS} lists identity and information of the 27 features.

\begin{figure}[h!]
\begin{center}
\includegraphics[width=9cm]{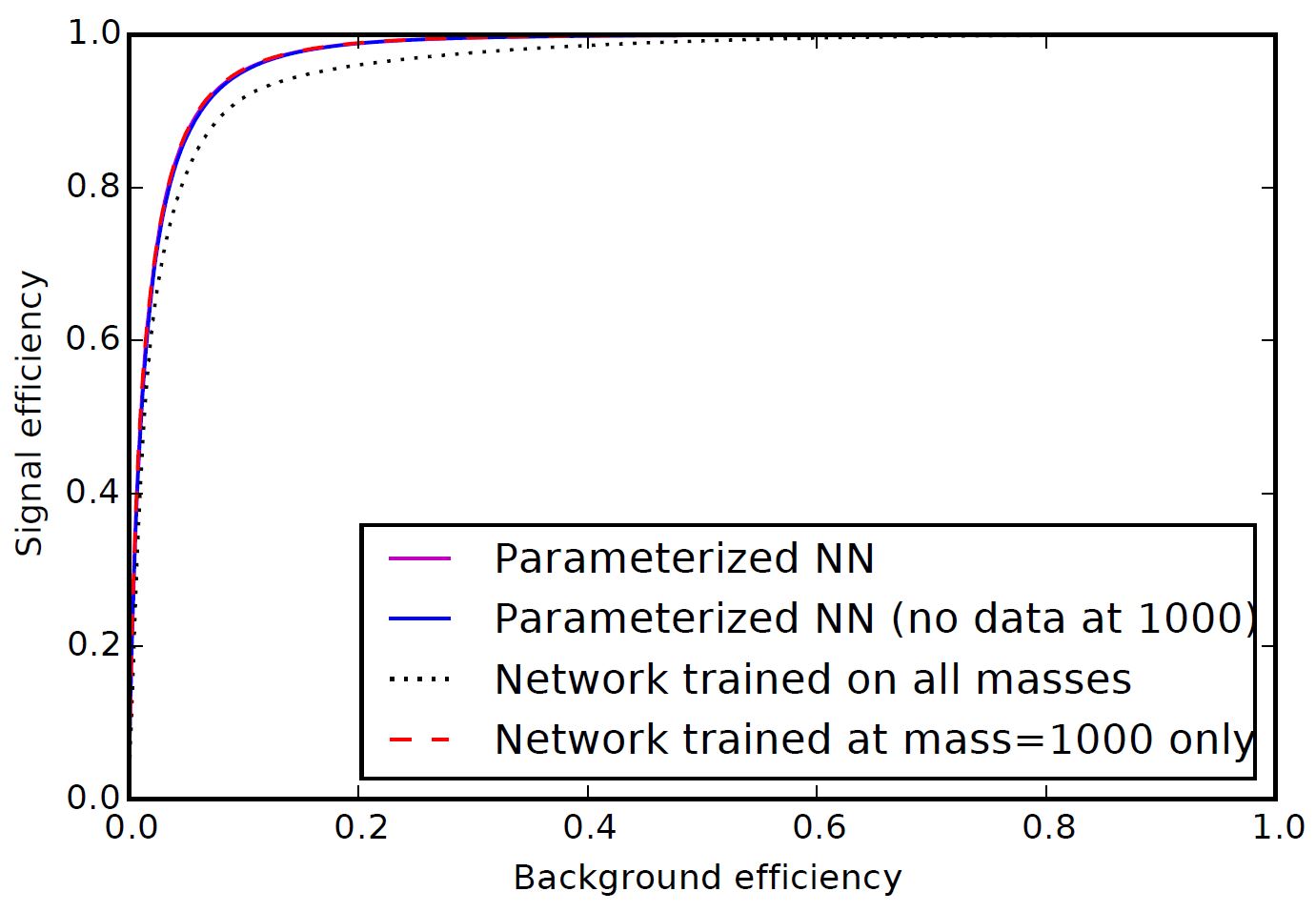}
\small
\caption{\small  Comparison of signal and background efficiency curves for four classes of neural networks on the HEPMASS dataset. Of relevance here is the dashed red curve, which correspond to a non-parametrized network trained and tested on the sample of reference, with a resonance mass of 1000 GeV. Reprinted with permission from~\cite{whiteson2}.}
\normalsize
\label{f:whiteson}
\end{center}
\end{figure}

\noindent
In~\cite{whiteson2} several ROC curves are presented to compare the performance of parametrized and non-parametrized neural networks on the HEPMASS signal discrimination problem. Those are the result of supervised classification, which duly exploits {\em a priori} knowledge of the signal density. As can be seen in Fig.~\ref{f:whiteson}, the non-mass-parametrized neural network achieves a background efficiency of about $3\%$ for a signal efficiency of $80\%$, {\em e.g.}. We will use these approximate values for a qualitative comparison to the performance of \texttt{RanBox}, bearing in mind all the caveats of any comparison of supervised and unsupervised classification methods.

In this section we use a mixture of signal and background events from the HEPMASS dataset to test under what conditions \texttt{RanBox} is capable of evidencing feature space regions with a dominant signal contamination. Since the feature space is rich with interdependencies among the features, the task of a unsupervised algorithm is considerably harder than in the case of the synthetic dataset studied in Sec.~\ref{s:synthetic}, as significant overdensities are expected to arise from the structure of background processes alone. Furthermore, in a real-life application of \texttt{RanBox}, the user would be unable to extract the distribution of the test statistic under the null hypothesis, as even slight differences between simulation and real data would distort the output. We consider therefore that in that case \texttt{RanBox} would be used by running it on real data as they come, without any pretense of assessing a significance level of the returned overdense regions or of studying the power of a selection criterion, but rather with the aim of focusing the attention of researchers on the combinations of features that exhibit interesting localized overdensities.

\begin{figure}
    \begin{center}
    \includegraphics[width=10cm]{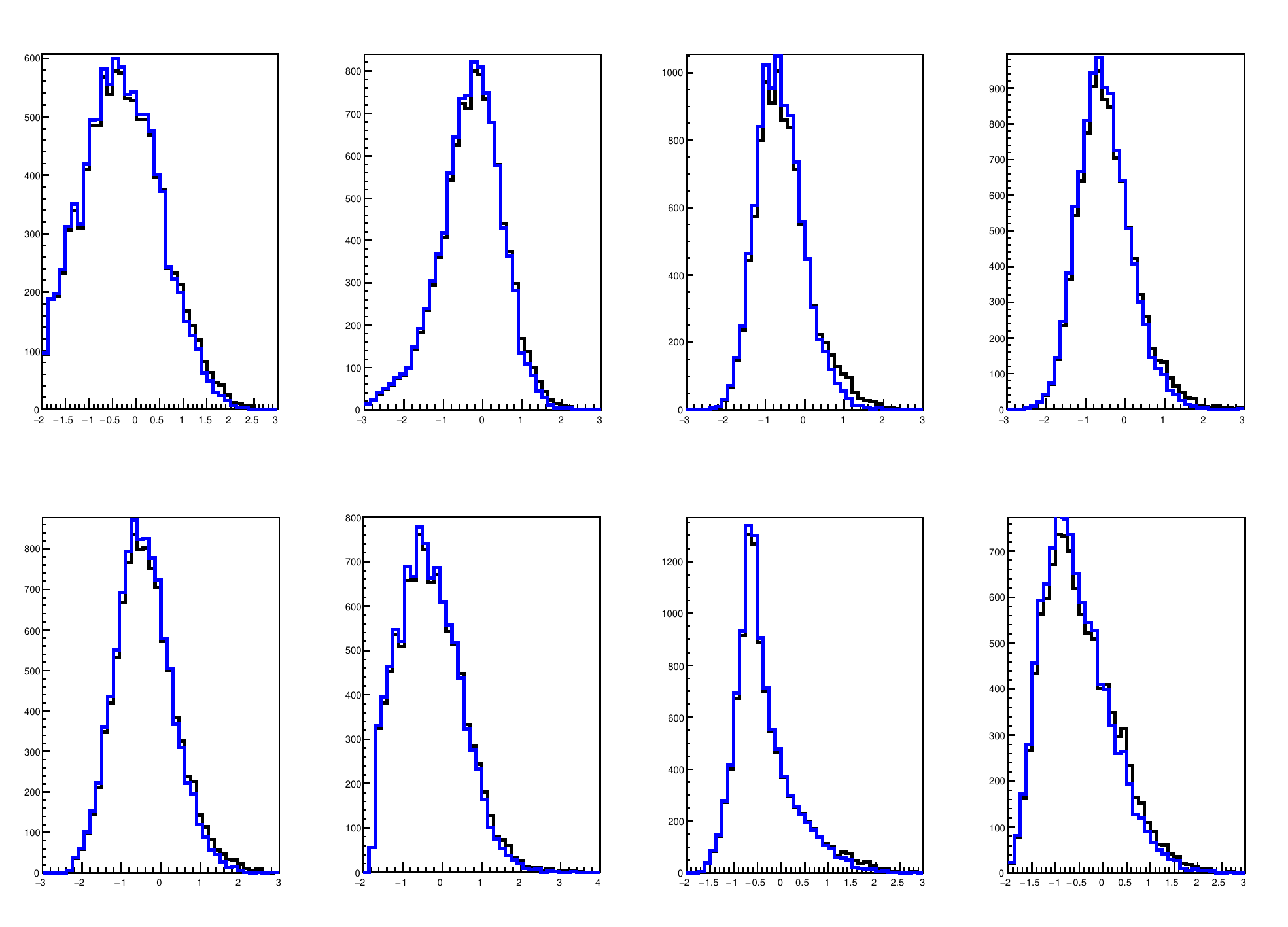}
    \caption{\small Comparison of the distribution of pure background (blue) and a mixture of 5\% signal and background (black) in the most discriminating features in the HEPMASS dataset. Left to right, top to bottom: features 0, 3, 6, 10, 14, 18, 25, and 26.}
\label{f:27features_mixture}
\end{center}
\end{figure}

\begin{table}[h!]
\begin{tabular}{l|lllllll}
Algorithm & Init. & T.S. & Extrap. & ${\cal{D}}'$ or ${\cal{D}}'_{max}$ & D red. & $N_{iter}$ & $N_{best}$ \\
\hline
\texttt{RanBox} & A2 & $R_1$ & SB & ${\cal{D}}'=12$ & no & 10,000 & N/A \\
\texttt{RanBoxIter} & A2 & $R_1$ & SB & ${\cal{D}}'_{max}=12$ & no & N/A & 50 \\
\hline
\end{tabular}
\small
\caption{\small Run parameters of the \texttt{RanBox} and \texttt{RanBoxIter} algorithms for a test on the HEPMASS dataset with a $5\%$ signal contamination. ``Init.'' indicates the method defining the initial dimension of the search box; ``Extrap.'' identifies the way by which a prediction of events in the box is computed; $\cal{D}'$ is the dimensionality of the subspaces scanned by \texttt{RanBox}, and ${\cal{D}}'_{max}$ the maximum dimension of the subspaces constructed iteratively by \texttt{RanBoxIter}; ``Dim. red.'' indicates whether the dimensionality of the feature space was reduced with PCA or by discarding the most correlated variables; $N_{iter}$ is the number of searched subspaces by \texttt{RanBox}; and $N_{best}$ is the number of feature combinations considered by \texttt{RanBoxIter} when adding one dimension to the previous subspaces search. }
\label{t:hepmasspars}
\normalsize
\end{table}

We proceed with exploratory runs of the \texttt{RanBox} and \texttt{RanBoxIter} algorithms on the HEPMASS dataset as we would perform them on real data. We construct a dataset comprised of 250 signal and 4750 background events: the $5\%$ signal fraction is small enough to make the signal indistinguishable in the marginal distributions of even the most discriminating variables, as shown in Fig.~\ref{f:27features_mixture}. We run \texttt{RanBox} and \texttt{RanBoxIter} with the parameters listed in Table~\ref{t:hepmasspars}. They constitute a reasonable choice for a run on HEPMASS. In particular, since we wish to be sensitive to a small signal contamination rather than having the algorithm get distracted by broader-scale background correlations, we initially consider that the $R_1$ test statistic might be more sensitive to a signal component. Also, we use the sidebands method to extrapolate the density in the search box, as this better factors out the local disuniformities in the data. The choice of dimensionality (or maximum dimensionality, for \texttt{RanBoxIter}) of the scanned subspaces is instead driven by preconceptions on the fact that a signal of new physics will most likely exhibit distinctive features only in a subset of the considered kinematical variables~\footnote{ Indeed, in the considered search for $X \to t \bar{t}$, apart from the resonant structure of the total invariant mass of the decay products, one expects only minor differences of the signal with respect to the non-resonant $t \bar{t}$ production predicted by the SM. }; 12 is anyway close to the maximum meaningful choice for that parameter, as is clear if we consider that in a 12-dimensional space a box of sides equal to half the range of each feature will on average contain only $5000\times 2^{-12}=1.2$ events out of 5000. Finally, we do not apply any dimensionality reduction to the input data, as we observe that the maximum two-variable correlation coefficient (0.757) in the mixture dataset is not particularly high. 

\begin{table}[h!]
\begin{tabular}{rrrrrrl}
\hline
$R_1$ & $N_{in}$ & $N_{exp}$ & $N_{s}$ & $\epsilon_s$ & Gain & Active features \\
\hline
52.78 & 54 & 0.02 & 46 & 0.184 & 17.04 & 101011100111010000000110001 \\
45.35 & 48 & 0.06 & 38 & 0.152 & 15.83 & 000111100011001011000100011 \\
41.60 & 46 & 0.11 & 33 & 0.132 & 14.35 & 100010010110111011000100001 \\
40.72 & 46 & 0.13 & 18 & 0.072 & 7.83  & 101000110110101000100010011 \\
40.38 & 44 & 0.09 & 41 & 0.164 & 18.64 & 100100100100010001110111001 \\
40.17 & 47 & 0.17 &  0 & 0.000 &  0.00 & 011001000100010111001100011 \\
39.82 & 44 & 0.10 &  0 & 0.000 &  0.00 & 100001010100011001010101011 \\
38.54 & 44 & 0.14 &  0 & 0.000 &  0.00 & 001001101101110001101100000 \\
38.36 & 44 & 0.15 & 30 & 0.120 & 13.91 & 000110101110010000001101011 \\
38.05 & 43 & 0.13 & 14 & 0.056 &  6.51 & 110000100110001110100011001 \\
\hline
\end{tabular}
\small
\caption{\small Results of an exploratory \texttt{RanBox} search on the HEPMASS dataset with a $5\%$ signal contamination; data for the 10 most significant boxes are reported. $N_s$ indicates the number of signal events in the search box; $\epsilon_s$ is the efficiency of the box selection for the signal component; gain is computed as the increase in the signal fraction of the box over the initial dataset.  For other detail see the text. }
\normalsize
\label{t:HEPMASSexpl1}
\end{table}

\begin{table}[h!]
\begin{tabular}{rrrrrrlr}
\hline
$R_1$ & $N_{in}$ & $N_{exp}$ & $N_{s}$ & $\epsilon_s$ & Gain & Active features & $\cal{D}'$\\
\hline
69.92 & 83 & 0.19 & 47 & 0.188 & 11.33 & 000000100110011001100011111 & 12\\
63.93 & 66 & 0.03 & 59 & 0.236 & 17.88 & 000101100110011000101100001 & 11\\
62.74 & 65 & 0.04 & 57 & 0.228 & 17.54 & 010000100100011001100111011 & 12\\
60.71 & 68 & 0.12 & 54 & 0.216 & 15.88 & 000000100100011011100111011 & 12\\
58.52 & 61 & 0.04 & 49 & 0.196 & 16.07 & 000100100110001001100101011 & 11\\
56.88 & 62 & 0.09 & 46 & 0.184 & 14.84 & 000000100110011000100001011 &  9\\
51.07 & 56 & 0.10 & 31 & 0.124 & 11.07 & 000000100110011000100111001 & 10\\
51.03 & 53 & 0.04 & 49 & 0.196 & 18.49 & 000100100100001001100101011 & 10\\
51.00 & 54 & 0.06 & 35 & 0.140 & 12.96 & 000100100110011010100100001 & 10\\
50.72 & 57 & 0.12 & 39 & 0.156 & 13.68 & 000010100110011000100001001 &  9\\
49.98 & 51 & 0.02 &  0 & 0.000 &  0.00 & 100001000110010001001100000 &  8\\
49.94 & 54 & 0.08 & 47 & 0.188 & 17.41 & 000000100100001000100111011 &  9\\
\hline
\end{tabular}
\small
\caption{\small  Results of an exploratory {RanBoxIter} search on the HEPMASS dataset with a $5\%$ signal contamination. The 12 most significant boxes are reported, such that we may include a negative result (the $11^{th}$ most significant box). ${\cal{D'}}$ is the dimensionality of the reported box. For other detail see the text. }
\normalsize
\label{t:HEPMASSexpl2}
\end{table}

% best box found by ranboxiter:
% var9 [0,0.68]
% var26 [0.78,1]
% var18 [0.17,1]
% var14 [0.66,1]
% var23 [0.75,1]
% var6 [0.75,1]
% var22 [0.11,1]
% var25 [0.52,1]
% var17 [0,0.79]
% var13 [0,0.2]
% var24 [0,1]
% var10 [0.79,1]

From the results listed in Table~\ref{t:HEPMASSexpl1} and~\ref{t:HEPMASSexpl2} we may draw a few interesting conclusions. First of all, the search of 10,000 subspaces performed by \texttt{RanBox} returns a good number of signal-rich regions, as five of the ten most significant boxes are dominated by the signal component, and two more are also considerably signal-enriched, by factors above six~\footnote{In the following we take that factor as a threshold to count the number of signal-rich (SR) boxes among the first ten boxes, a number we report as $SR_{1:10}$.}. Such an output, and in particular the most significant box alone, would certainly allow experimentalists to focus on the small signal now evident in the identified regions, hence we consider this output a success of the anomaly detection task. We also note that the scan of 10,000 12-dimensional subspaces costs nearly 10 hours of running on a single CPU; the scan of all 12-dimensional subspaces of the 27-dimensional feature space is instead not an easily viable option, as this would require $27!/(12!15!)=1.738 \times 10^7$ iterations, or about two years of CPU on a single machine. Regardless, on the HEPMASS dataset a limited number of combinations of 12 features still allow to evidence a small signal.

% GS: "by factors above 6" suggest replacing "6" with "six" due to footnote number --> ok

\texttt{RanBoxIter} is quicker to run with the above settings, as it completes its scan in little more than five hours of CPU, keeping a list of the 50 most significant boxes during its iterative scan (a parameter which scales roughly linearly with CPU time). It also returns, as the 10 most significant regions, ones which are all quite rich in signal. In Table~\ref{t:HEPMASSexpl2} we observe that the list of features defining the subspaces where the most significant boxes are found include a few ({\em e.g.} features 9, 18, and 26)
which are always used ---an indication that their combination is highly discriminant for the anomaly originating the overdense regions, and an observation that may be useful in the interpretation of results. One may also notice (see the rightmost column in Table~\ref{t:HEPMASSexpl2}) how in some cases the dimensionality of the most significant box is smaller than the maximum investigated in the run (12), which shows that the algorithm cannot always fruitfully exploit the value of ${\cal{D}}'_{max}$ to maximize the test statistic. In general this might indicate that the signal does not possess discriminating characteristics in enough of the features, but in the case at hand it rather reflects the relatively small number of events of the data sample over which the algorithm is run: the identified boxes have sidebands already devoid of any background event, hence the maximization of the test statistic does not favour further losses of events in the signal box by cuts on additional features~\footnote{As mentioned {\em supra} (Sec.~\ref{s:teststatistic}), for zero events in the sideband region the estimate $N_{in, \tau}$ reverts to $N_{in, V}$.}.

By comparing the results shown in the two tables above, one may conclude that \texttt{RanBoxIter} performs better than \texttt{RanBox} on this particular application, as it returns signal-rich regions in all its ten most significant boxes, and the signal-to-noise gain resulting in those regions is higher than those of the best regions reported by \texttt{RanBox}. This result proves that the incremental scan is effective, especially in situations such as the one considered here, where the large combinatorial factor that \texttt{RanBox} has to overcome in order to examine combinations of features sensitive to a small signal prevents it from reaching its full sensitivity.

If we now compute the signal and background efficiency of the regions returned by the algorithms in their exploration of the $f_s=5\%$ datasets, we notice that the best box identified by \texttt{RanBox} contains 46 signal events out of 54, which corresponds to an 85\% efficiency; the background efficiency is instead 8/4950 = 0.16\%. For \texttt{RanBoxIter} the signal and background efficiencies of the best box are instead respectively 94\% and 0.73\%. In both cases these numbers compare quite favourably to those of the neural network results graphically displayed in Fig.~\ref{f:whiteson}. We stress again the modest value of this observation, given the improper nature of a comparison of this kind. In particular, the \texttt{RanBox} and \texttt{RanBoxIter} results have unknown generalization properties ---they are obtained from a single dataset, on which multiple testing is performed: the performance would be less good on a different testing sample. On the other hand, the search algorithms were only shown a total data sample of 5000 events, a number over two orders of magnitude smaller than the training sample of the neural networks.  

%In summary, both algorithms prove useful in the considered application, and may provide analysts with a new way to scan large parameter spaces of event collisions.

%%%%%%%%%%%%%%%%%%%%%%%%%%%%%%%%%
\subsubsection {Further studies}

\noindent
The above tests prove the usefulness of the algorithm in both its instantiations, but leave open the question of what are their limits of sensitivity to a smaller signal contamination. In Table~\ref{t:HEPMASSresults_iter} we therefore present the results of a few additional runs of \texttt{RanBoxIter} on sets of 5000 events, a variable fraction of which are taken from the HEPMASS signal simulation and the rest from the corresponding background simulation. In the reported tests we fix the maximum dimensionality of the iterative scans to ${\cal{D}}'_{max}=12$, use Algorithm 1 or Algorithm 2 for initialization of the search box, and employ the $R_1$ test statistic in the maximization search. We operate no dimensionality reduction technique to the input data. We provide some commentary of the results below.\par

\small
\begin{table}[h!]
\begin{center}
    \begin{tabular}{rrrrrrrrr}
    \hline
    Test & $N_s/N_b$ & $R_1$ & $N_{in}$ & $ N_{exp}$ & $N_{s}$ & Gain  & $SR_{1:10}$ & $\overline{\epsilon_s^{1:10}}$ \\ 
    \hline
    1    & 200/4800  & 51.12    & 55    & 0.075  & 48  & 21.82 &  9 & 0.164 \\
    2    & 150/4850  & 48.61    & 52    & 0.069  &  0  &  0.00 &  4 & 0.037 \\
    3    & 100/4900  & 41.20    & 44    & 0.067  &  0  &  0.00 &  2 & 0.024 \\
    4    &  80/4920  & 44.48    & 47    & 0.056  &  8  & 10.64 &  9 & 0.112 \\ 
    5    &  70/4930  & 52.45    & 56    & 0.067  & 14  & 17.86 &  7 & 0.113 \\ 
    6    &  60/4940  & 46.74    & 51    & 0.091  &  1  &  1.64 &  2 & 0.027 \\
    7    &  50/4950  & 42.16    & 43    & 0.019  &  0  &  0.00 &  3 & 0.028 \\
    \hline
    8    & 200/4800  & 33.05    & 35    & 0.058  & 32  & 22.86 &  6 & 0.098 \\
    9    & 150/4850  & 38.24    & 40    & 0.045  & 31  & 25.83 & 10 & 0.175 \\
    10   & 100/4900  & 31.50    & 63    & 1.000  &  0  &  0.00 &  1 & 0.001 \\
    11   &  80/4920  & 31.66    & 33    & 0.042  &  0  &  0.00 &  5 & 0.065 \\ 
    12   &  70/4930  & 41.50    & 83    & 1.000  &  0  &  0.00 &  1 & 0.013 \\ 
    13   &  60/4940  & 36.47    & 37    & 0.014  &  0  &  0.00 &  7 & 0.088 \\
    14   &  50/4950  & 45.50    & 91    & 1.000  &  5  &  5.49 &  4 & 0.066 \\
    \hline
    \end{tabular}
\small
\caption{\small Results of a few additional \texttt{RanBoxIter} test runs on 5000 events from the HEPMASS dataset, using Algorithm 2 (tests 1-7) and Algorithm 1 (tests 8-14) for the initialization of the search boxes, and scans of subspaces with dimension up to ${\cal{D}}'_{max}=12$. The last two columns report the number of signal-rich boxes and the average signal efficiency of the ten most significant boxes. See the text for more details.}
\normalsize
\label{t:HEPMASSresults_iter}
\end{center}
\end{table}
\normalsize

\begin{figure}[h!]
\begin{center}
\includegraphics[width=14cm]{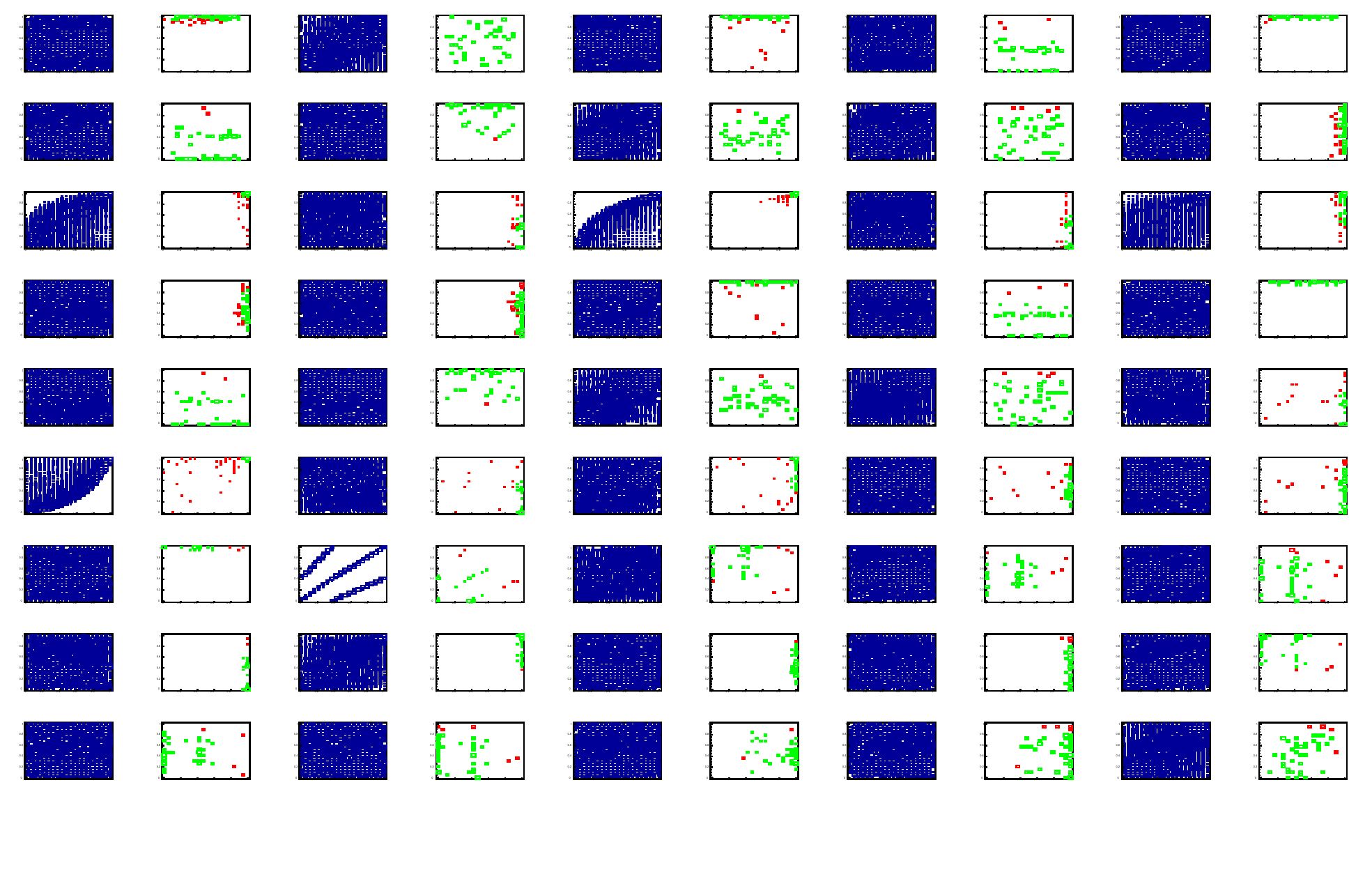} % test 9 table 5
\caption{\small Two-feature copula-space distributions from test 9 in Table~\ref{t:HEPMASSresults_iter}. Each left graph in a pair (columns 1,3,5,7,9) shows the totality of the data (blue scatterplots with 5000 events, including a 3\% signal component) on the 45 combinations of the 10 features (out of ${\cal{D}}'_{max}=12$) defining the highest-$R_1$ box; the corresponding graph on the right shows the data selected in the best box (in green), and data failing the selection only because of the values of the features shown in the graph (in red). From left to right, the shown pairs of features are:  [1,6],[1,11],[1,14],[1,9],[1,10] (row 1); [1,13],[1,25],[1,7],[1,19],[6,11] (row 2); [6,14],[6,9],[6,10],[6,13],[6,25] (row 3);
[6,7],[6,19],[11,14],[11,9],[11,10] (row 4);
[11,13],[11,25],[11,7],[11,19],[14,9] (row 5);
[14,10],[14,13],[14,25],[14,7],[14,19] (row 6);
[9,10],[9,13],[9,25],[9,7],[9,19] (row 7);
[10,13],[10,25],[10,7],[10,19],[13,25] (row 8);
[13,7],[13,19],[25,7],[25,19],[7,19] (row 9).}
\label{f:test9table5}
\end{center}
\end{figure}

The tests show that \texttt{RanBoxIter} has no difficulty in spotting a 2\% to 4\% signal in a HEPMASS sample of 5000 events, when Algorithm 2 is used for the box initialization. For tests 1 to 7, the decreasing value of the number of signal-rich regions and signal efficiency in the 10 most significant boxes (last two columns in the table) with decreasing signal contamination indicates that 2\% is indeed close to the limit when local background fluctuations or overdense region start to compete and offset the small signal. Tests 8 to 14 further show that the initialization provided by Algorithm 1 is equally or maybe slightly less effective on this dataset, as searches converge to boxes with smaller values of $R_1$, with a minor reduction in sensitivity. Figure~\ref{f:test9table5} shows the two-dimensional distributions of the subspace identified by test 9.

\begin{table}[h!]
\begin{center}
    \begin{tabular}{rrrrrrrr}
    \hline
    Test & $N_s/N_b$ & T.S. max & $N_{in}$/$N_{exp}$ & $N_{s}$ & Gain & $SR_{1:10}$ & $\overline{\epsilon_s^{1:10}}$ \\
    \hline
    1 & 250/4750       & $Z_{PL}=28.06$     &   39/0.00  & 35      & 17.95  & 8 & 0.097\\
    2 & 200/4800       & $Z_{PL}=27.46$     & 1003/133.00&  0      &  0.00  & 6 & 0.079\\
    3 & 150/4850       & $Z_{PL}=24.45$     &   24/0.00  & 21      & 29.17  & 7 & 0.140\\
    4 & 100/4900       & $Z_{PL}=26.95$     &   45/0.01  &  0      &  0.00  & 1 & 0.014\\
    5 &  80/4920       & $Z_{PL}=24.65$     &   43/0.05  & 14      & 20.35  & 3 & 0.044\\
    6 &  70/4930       & $Z_{PL}=23.56$     &   41/0.01  &  0      &  0.00  & 1 & 0.010\\
%    7 & 250/4750       & $R_{1}=57.57$      &   63/0.09  & 54      & 17.14  & 7 & 0.102\\
\hline
    7 & 250/4750       & $R_{1}=52.78$      &   54/0.02  & 46      & 17.04  & 7 & 0.092\\
    8 & 200/4800       & $R_{1}=50.78$      &   60/0.18  & 33      & 13.75  & 6 & 0.097\\
    9 & 150/4850       & $R_{1}=43.58$      &   53/0.21  & 15      &  9.44  & 6 & 0.071\\
    10& 100/4900       & $R_{1}=45.29$      &   49/0.08  &  0      &  0.00  & 3 & 0.038\\
    11&  80/4920       & $R_{1}=51.22$      &   52/0.02  &  0      &  0.00  & 0 & 0.000\\
    12&  70/4930       & $R_{1}=43.44$      &   48/0.10  &  0      &  0.00  & 0 & 0.000\\
    \hline
    \end{tabular}
\caption{\small Sample results of \texttt{RanBox} runs on 5000 events from the HEPMASS dataset, with varying signal fraction and the two choices of test statistic. See the text for more details.}
\label{t:HEPMASSresults}
\end{center}
\end{table}

We perform a similar investigation using \texttt{RanBox}, as detailed in Table~\ref{t:HEPMASSresults}, using 10,000 trials for the subspace sampling and a subspace dimensionality of ${\cal{D}}'=12$. This time we start (see test 1) by searching for a 5\% signal in a set of 5000 events using the $Z_{PL}$ test statistic. The algorithm returns as the most significant box one which is rich in signal component, and we observe that the three next-best-significance boxes (not reported in Table~\ref{t:HEPMASSresults}) are similarly enriched in signal events. We gradually reduce the signal fraction in tests 2-6 and observe that results are not uniform: \texttt{RanBox} in some cases identifies as the most significant box one devoid of signal. In general we observe that the number of boxes that are signal-enriched among the first 10 ($SR_{1:10}$) usually decreases as initial signal fraction is reduced; the average signal efficiency also becomes smaller. Yet the algorithm finds significantly signal-enriched boxes among the first 10 even for an initial signal fraction of 1.4\% and 1.6\%. We also observe that in test 2 the $Z_{PL}$ maximization focuses on a very wide box, an indication of the existence of broad-scale multivariate density variations of the background component of this dataset.

Based on the above observation, in tests 7-12 we turn our attention to the $R_1$ test statistic, which should give more importance to smaller feature-space regions. This indeed allows \texttt{RanBox} to converge on signal-rich regions when the signal fraction of the data sample is 3\% or larger; for smaller signal fractions, however, \texttt{RanBox} becomes unable to evidence the signal component in the reported overdense regions.

In all the tests reported above we stuck to data samples of 5000 events. The reason of that choice is that \texttt{RanBox} and \texttt{RanBoxIter} in their present implementation have not been designed to handle much larger datasets, as their CPU consumption is significant, especially in the initial k-NN (Algorithm 1) or kernel-based (Algorithm 2) initialization of the search box boundaries. %To investigate the performance of the search method on a larger sample of data, we leverage the identification of a signal-sensitive subspace of the feature space provided by the run described in Table \ref{t:HEPMASSexpl1}, and only scan the parameter space defined by features 0,2,4,5,6,9,10,11,13,21,and 26, which provided the highest-$R_1$ box among 10,000 tested. We use for these tests a total data sample of 50,000 events, wherein we vary the fraction of signal events. Since the absolute number of these is higher by a factor of 10 with respect to previous tests, for a given signal fraction, we expect that we will be able to push the sensitivity to smaller signal fractions, as the $R_1$ test statistic will be a more robust estimator of the density variations in the scanned space.
In order to test the performance of the search method over a larger sample of HEPMASS data we use \texttt{RanBoxIter}, which on that dataset requires less CPU to produce stable results, and we further lessen the CPU load by using Algorithm 0 and reducing $N_{best}$ to 20, accepting the slight loss of performance that these settings offer. We use a total of 100,000 events, wherein we mix fractions of signal events smaller than those those which allowed the previously reported runs to converge to signal-rich boxes. Since for any given signal fraction the absolute number of signal events is now higher by a factor of 20 with respect to all previous tests, we expect to gain some more sensitivity to smaller signal fractions, as in the higher-statistics conditions the $R_1$ test statistic is a more robust estimator of density variations in the scanned space.

\begin{table}[h!]
\begin{center}
    \begin{tabular}{rrrrrrrr}
    \hline
    Test & $N_s/N_b$ & $R_1$ & $N_{in}$/$N_{exp}$ & $N_{s}$ & Gain & $SR_{1:10}$ & $\overline{\epsilon_s^{1:10}}$ \\
    \hline
    %1 &  500/49,500 & & & & & & \\
    %2 &  250/49,750 & 46.50 &  29/0.008 & 27 & 186.2 & &  \\ % Nbest=1, redo with Nbest 20
    %3 &  200/49,800 & 36.98 &  37/0.000 &  0 &   0.0 & 1 & 0.006  \\
    %4 &  150/49,850 & 41.46 &  42/0.012 &  0 &   0.0 & 1 & 0.026  \\
    1 & 1000/99,000 & 40.94 &  41/0.001 & 35 &  85.4 & 3 & 0.009  \\ %nbest 20, a0
    %6 &  900/99,100 & & & & & \\ % Nbest 20
    2 &  800/99,200 & 46.10 &  52/0.125 & 46 & 110.6 & 5 & 0.025  \\
    %8 &  700/99,300 & & & & & \\
    3 &  600/99,400 & 42.00 &  84/1.000 &  0 &   0.0 & 2 & 0.008  \\
    4 &  500/99,500 & 37.00 & 185/4.000 &  0 &   0.0 & 4 & 0.012  \\ % Nbest 20
    5 &  400/99,600 & 51.00 & 102/1.000 &  0 &   0.0 & 1 & 0.001  \\
    6 &  300/99,700 & 36.96 &  37/0.001 &  0 &   0.0 & 1 & 0.007  \\
    \hline
    \end{tabular}
\caption{\small Sample results of \texttt{RanBoxIter} runs on 100,000 event samples from the HEPMASS dataset, with varying signal fraction. See the text for more details.}
\label{t:HEPMASShighstat}
\end{center}
\end{table}

\noindent
As shown in Table \ref{t:HEPMASShighstat}, indeed \texttt{RanBoxIter} is capable of identifying signals contaminating the data samples with fractions smaller than 1\% if in absolute terms the signal is large enough to produce significant overdensities. We further observe that if the users do not limit themselves to examining the highest-$R_1$ box and verify the distributions of data in the 10 best boxes, they will be able to observe anomalous regions contributed by significant signal fractions in datasets where the signal contribution is as small as 0.5\% or less. For example, the single signal-rich box among the 10 best boxes found in test 6, where the initial signal fraction is of 0.3\%, contains 28 events (with zero expected), of which 19 are from the signal component; the corresponding gain in signal fraction is a factor of 226. 

We believe the above observations prove the usefulness of \texttt{RanBoxIter} scans of the typical datasets studied in high-energy physics searches for new physics, as an exploratory {\em modus operandi} involving the systematic examination of anomalous regions would allow experimenters to spot regions of phase space that constitute valid starting points of broader investigations. They would {\em e.g.} be able to compare the observed distributions to ones predicted by SM simulations, as well as define blind selection strategies for 
analysis of data collected in future runs of the collider.

The results also allow us to draw some conclusions on the most performing settings of \texttt{RanBox} to be used in the HEPMASS use case. Here, however, we stress one important point: by telling the tale of how these choices may be defined based on sample test results, we are implicitly declaring how the algorithm ---but in general, we believe, any unsupervised search--- requires an {\em ad hoc} tuning to perform its task most effectively. This is not to be taken as a demonstration that this kind of search is useless: quite on the contrary, the tool can be a very useful one in examining the properties of multi-dimensional data. It cannot, on the other hand, be employed as a catch-all machine ready to identify an anomaly in an arbitrary dataset: this is nothing else than a by-product of the well-known absence of a universal high-power test statistic, when the alternative hypothesis is not specified.

%%%%%%%%%%%%%%%%%%%%%%%%%%%%%%%%%%%%%%%%%%%%%%%%%%%%%%%
\subsection{Electron neutrino identification in MiniBooNE \label{s:experiments_miniboone}}

\noindent
For a second research application wherein to test the performance of \texttt{RanBox} and \texttt{RanBoxIter}, we turn to another public dataset available in the UCI ML repository and those which possess the characteristics that are most suitable to our algorithm. In particular, we look for a dataset suitable for binary classification where the majority of features are non-categorical, with a total number of features not exceeding several tens, and at least several thousand available examples for both classes.

\begin{figure}[h!]
\begin{center}
\includegraphics[width=12cm]{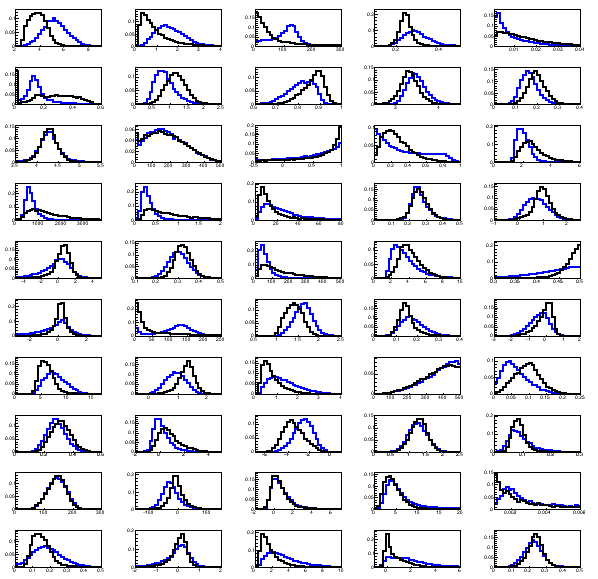}
\caption{\small Normalized and standardized distributions of the 50 features of electron-neutrino candidate interactions in MiniBooNE. Distributions of real electron interactions are in black, and backgrounds are in blue. }
\label{f:miniboonefeatures}    
\end{center}
\end{figure}

The \texttt{MiniBooNE} dataset~\cite{minibooneUCI} suits the above requirements. \texttt{MiniBooNE}~\cite{miniboone} is a neutrino experiment built to study neutrino oscillations at Fermilab, in particular in order to shed further light in the so-called LSND anomaly~\cite{lsnd}. The public dataset available in the UCI repository contains about 130,000 events with 50 features per event (see Fig.~\ref{f:miniboonefeatures}). The classification task consists in separating real electron neutrino appearance events from spurious backgrounds where the event is not due to a charged-current electron neutrino interaction in the detector. The dataset was used in pioneering studies of the application of boosting techniques to decision tree classification~\cite{roe,roe2}. 

Given the dimensionality of the feature space, this dataset lends itself to an investigation of dimensionality-reduction techniques at a data preprocessing stage, in order to reduce the CPU demand of the subspace sampling. We will therefore study the effect of reducing to 40 or 30 the feature space dimensionality by selecting the corresponding highest-variance components with PCA, or alternatively select by CVR the features which present collectively the smallest amount of inter-correlation, as described {\em supra} (see Sec.~\ref{s:preprocessing}). We again start with a $(5\%:95\%)$ signal:background mixture sample of 5000 events, and perform tests with the two versions of the algorithm. We employ $R_1$ as the test statistic, Algorithm 2 for initialization, and the SB method to compute the expectation value of events in the boxes. The (maximum) dimensionality of the subspace scans is set to ${\cal{D}}' ({\cal{D}}'_{max})=12$. \texttt{RanBox} is run on 10,000 subspaces, and \texttt{RanBoxIter} keeps track of the 50 most significant boxes as it scans iteratively higher-dimensional subspaces.

As shown in Table~\ref{t:ranbox_miniboone} (test 1), \texttt{RanBox} returns as the most significant identified 12-dimensional box one considerably enriched in the small signal component; among the ten best boxes returned by the algorithm, only one has a signal fraction increased below a factor of six.
%~\footnote{For the purpose of reporting the number of signal-rich boxes $SR_{1:10}$ among the 10 highest-$R_1$ ones, we count those where the signal fraction is above this threshold.}. 
This is a demonstration that on this dataset the signal can be extracted even by scanning a very small fraction of the available subspaces (10,000 out of $1.21 \times 10^{11}$). The most significant box identified by this run is one containing 49 signal events out of 69, which correspond to 19.6\% of the total signal present in the data sample, and 20 background events, which correspond to a background efficiency of 0.42\%~\footnote{Comparing these figures to the original results of~\cite{roe,roe2} looks odd not only because of the improper nature of the comparison of supervised and unsupervised learning, but also because the latter are 15 years old at the time of writing; we abstain from doing so here. }.
% GS: Again suggest changing "6" to "six" due to footnote number
Tests 2 and 3 instead show that PCA worsens the performance of \texttt{RanBox} in this data sample, removing sensitivity to a 5\% signal component; while CVR (tests 4 and 5) appears to increase it or anyway maintain it intact. Of course, given the smallness of the fraction of investigated subspaces (10,000) in all tests, these results are affected by large statistical fluctuations, and they can only be taken as a non-conclusive indication of the merits of the two techniques.

Similar insight is offered by examining the results of iterative scans of \texttt{RanBoxIter} (tests 6-10). We again note that on this particular data sample the reduction of dimensionality operated by PCA worsens the performance of the algorithm, which are instead uncompromised by using the CVR technique. Concerning the latter, in the case at hand the removal of the 10 most correlated features results in a maximum correlation of 0.679 among the remaining 40, while the removal of twenty features leaves a maximum correlation of 0.549; the original maximum correlation observed among the 2450 pairs of features was 0.910. 

%\small
\begin{table}[h!]
\begin{center}
    \begin{tabular}{rlrrrrrr}
    \hline
    Test & $\cal{D}$ & $R_1$ & $N_{in}$ & $N_{exp}$ & $N_{s}$ & $SR_{1:10}$ & $\overline{\epsilon_s^{1:10}}$ \\ % & Gain  & $\overline{G^{1:10}}$ & $\overline{f_s^{1:10}}$ \\ 
    \hline
    %1 & 50 &  72.87 & 73 & 0.00 & 39 & 9 & \\ % 10.68 & 11.71 & 0.1620 \\
    1 & 50       &  67.3  & 69 & 0.03 & 49 & 8 & 0.139 \\
    2 & 40 (PCA) &  36.5  & 73 & 1.00 &  2 & 0 & 0.007 \\ % pca10   
    3 & 30 (PCA) &  37.5  & 75 & 1.00 &  0 & 1 & 0.014 \\ % pca20
    4 & 40 (CVR) &  90.44 & 98 & 0.08 & 62 & 8 & 0.146 \\ % cvr10
    5 & 30 (CVR) &  64.63 & 68 & 0.05 & 42 & 10& 0.161 \\ % cvr20
    \hline
    6 & 50       & 101.77 & 106& 0.04 & 58 &  10 & 0.215 \\ %10.94 &     11.51 & 0.2148 \\
    7 & 40 (PCA) &  31.0  &  62& 1.00 &  0 &   0&  0.001 \\ % pca10
    8 & 30 (PCA) &  30.5  &  61& 1.00 &  0 &   0 & 0.001 \\ % pca20
    9 & 40 (CVR) & 103.85 & 111& 0.07 &111 &  10 & 0.208\\
%    9 & 40 (CVR) &  85.6  &  86& 0.00 & 52 &  10 & 0.234 \\ % cvr10
    10& 30 (CVR) &  52.8  &  57& 0.08 & 32 &  10 & 0.120 \\ % cvr20
    \hline
\end{tabular}
\caption{\small Results of tests on the MiniBooNE data sample composed of 250 signal and 4750 background events; tests 1-5 are performed with \texttt{RanBox} and tests 6-10 with \texttt{RanBoxIter}; see the text for other detail. }
\label{t:ranbox_miniboone}
\end{center}
\end{table}

\begin{figure}[h!] 
\begin{center}
    \includegraphics[width=14cm]{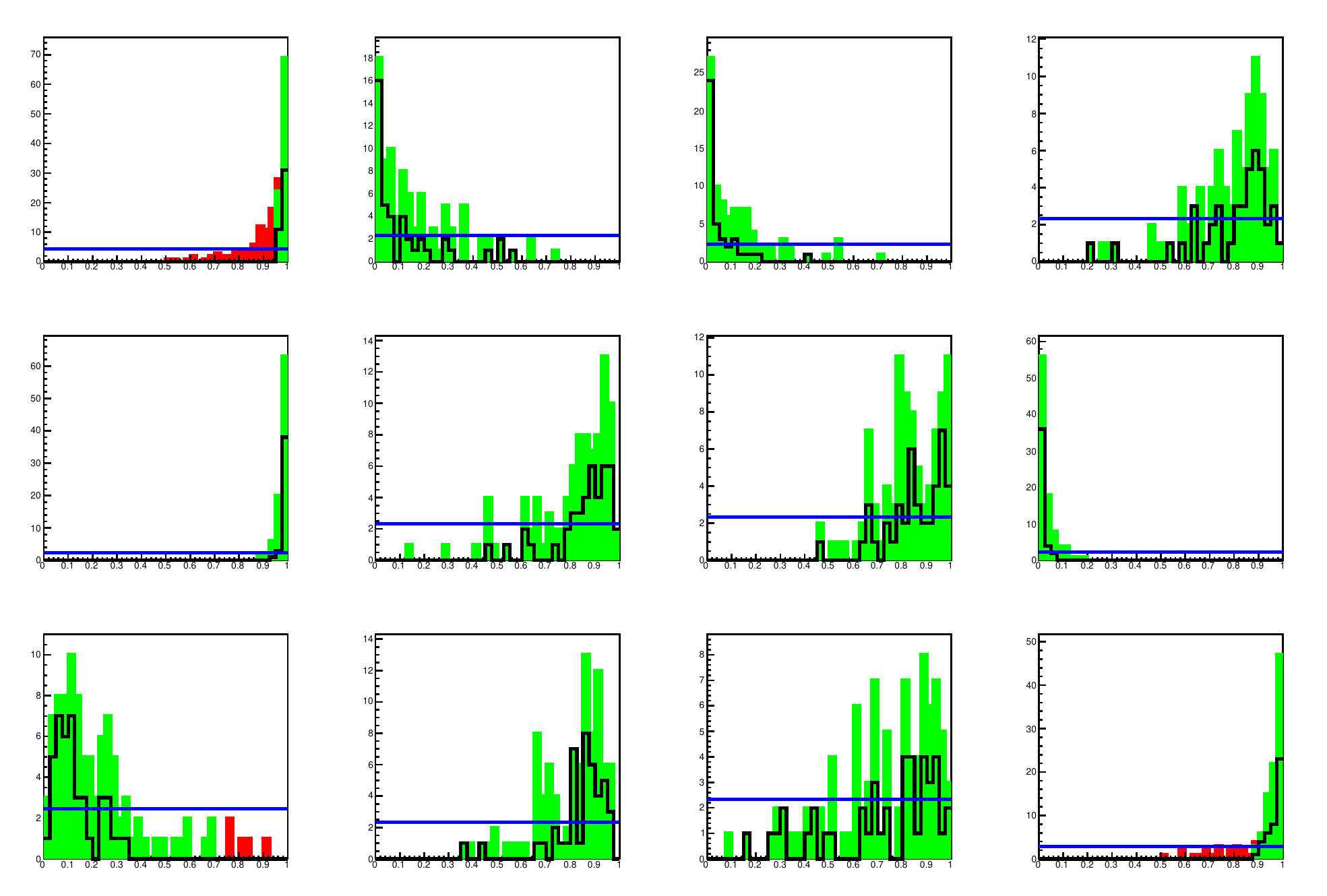}
    \caption{\small  Marginal distributions in copula space for features defining the best box found by \texttt{RanBoxIter} on MiniBooNE data for test 14 in Table~\ref{t:ranbox_miniboone}: the 12 features defining the subspace are shown for all data (blue empty histogram), for data in the best box (green), for data failing only a selection on the variable displayed (red), and for signal events in the box (black empty histogram with thick line). All distributions are normalized to unit area.}
    \label{f:test3miniboone}
\end{center}
\end{figure}

\begin{table}[h!]
\begin{center}
    \begin{tabular}{rrrrrrrr}
    \hline
    Test & $N_s/N_b$ & $R_1$ & $N_{in}$ & $N_{exp}$ & $N_{s}$ & $SR_{1:10}$ & $\overline{\epsilon_s^{1:10}}$ \\ 
    \hline
    9 & 250/4750 &103.85 & 111 & 0.068&111& 10& 0.208\\
    11& 225/4775 & 95.90 & 106 & 0.105& 62& 10& 0.183\\
    12& 200/4800 & 85.65 &  86 & 0.003& 52& 10& 0.234\\
    13& 175/4825 & 77.33 &  81 & 0.047&  4&  4& 0.053\\
    14& 150/4850 & 89.50 &  93 & 0.039& 42& 10& 0.368\\
    15& 125/4875 & 61.50 &  67 & 0.088& 31&  9& 0.168\\
    16& 100/4900 & 60.73 &  61 & 0.004& 27& 10& 0.195\\
    17&  80/4920 & 70.40 &  71 & 0.008& 23& 10& 0.205\\
    18&  70/4930 & 79.15 &  85 & 0.074& 21&  7& 0.160\\
    19&  60/4940 & 70.60 &  74 & 0.047&  0&  1& 0.030\\
    20&  50/4950 & 67.50 &  76 & 0.124&  1&  8& 0.186\\
    \hline
    \end{tabular}
    \caption{\small Tests of sensitivity of \texttt{RanBoxIter} to varying signal fraction in MiniBooNE data. All tests use CVR preprocessing of the data samples to remove the 10 most correlated features. See the text for other detail.}
\label{t:sf_miniboone}
\end{center}
\end{table}

A note we need to make, when comparing the average gain in signal to background ratio and average signal fraction of the ten best boxes returned by \texttt{RanBox} and \texttt{RanBoxIter}, is that the latter algorithm returns overdense regions in subspaces which share most of the features, because of the way it is designed. On the contrary, \texttt{RanBox} finds totally independent solutions to the problem, and thus is bound to return boxes whose signal-like characteristics are less homogeneous.

Having observed that CVR is useful with MiniBooNE data, we fix the number of removed features to 10 in our further tests, which are aimed at quantifing the minimum signal fraction to which the algorithm is sensitive  by progressively reducing its value. We use \texttt{RanBoxIter} for this study, as its results do not depend on a stochastic choice of features in the examined small fraction of subspaces~\footnote{ The trade-off, as already noted, is the dependence of \texttt{RanBoxIter} on fluctuations in the low-dimensional densities: {\em e.g.}, only 50 combinations out of 40x39=1560 are used as two-dimensional seeds for larger-dimension scans, and similar reductions occur every time the dimensionality is increased.}. Apart from changes in the signal fraction, all run parameters are kept equal to those of test 9 of Table~\ref{t:ranbox_miniboone}.
Figure~\ref{f:test3miniboone} shows the nine marginal distributions of the best box returned by \texttt{RanBoxIter} in test 14 of Table~\ref{t:sf_miniboone}, where a conspicuous signal is identified. 

From the tests reported in Table~\ref{t:sf_miniboone}, we observe that \texttt{RanBoxIter} is sensitive to signal fractions down to 1.5\% in MiniBooNE data with the settings we employed. It is of course likely that different choices for initialization algorithm, number of studied boxes $N_{best}$, and maximum dimensionality, along with higher statistic of the studied data sample would reduce that fraction still further, but an optimization of those parameters is too dataset-specific to be interesting, and is anyway beyond the scope of this work.

%%%%%%%%%%%%%%%%%%%%%%%%%%%%%%%%%%%%%%%%
\subsection{Credit card fraud detection}

\noindent
We finally turn to a non-research application of \texttt{RanBox}, by considering a dataset of online credit card transactions by European cardholders~\cite{creditcarddata}. 
% https://www.kaggle.com/mlg-ulb/creditcardfraud 
The data have a structure that suits the characteristics of our anomaly detector, as it is a tall dataset of 284,807 examples, with 30 non-categorical features. The original features include the time of the transaction, the amount, and a few numerical variables resulting from a PCA transformation ---hence we will abstain from applying again that procedure in our data preprocessing step.

\begin{figure}[h!]
\begin{center}
\includegraphics[width=12cm]{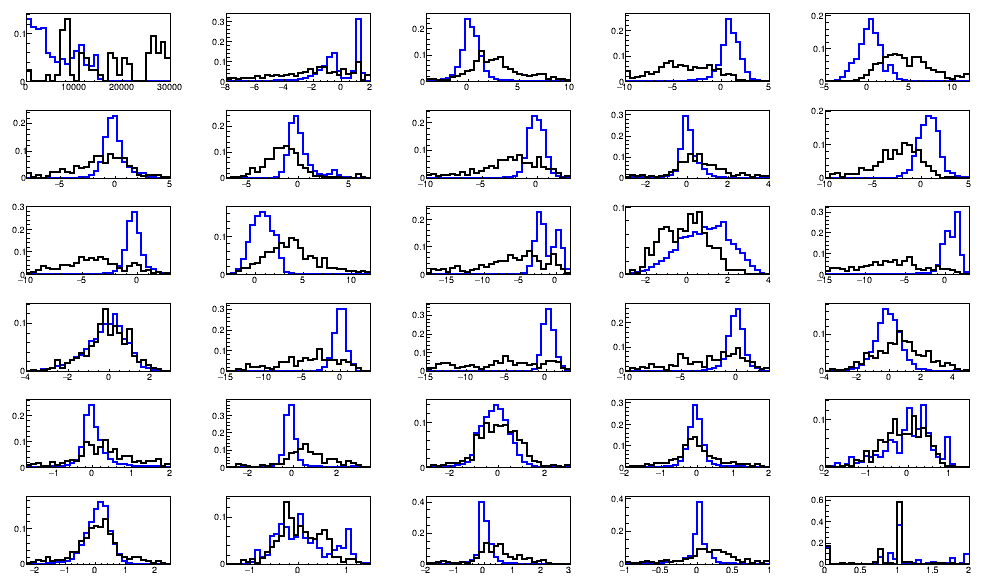}
\caption{ \small Normalized and standardized distributions of the 30 features of credit card transactions. Distributions of fraudulent transactions are shown in black, and regular ones are shown in blue. }
\label{f:fraudfeatures}    
\end{center}
\end{figure}

\noindent
Fraudulent transactions, which we will consider our signal in the following, constitute a very small fraction of the original dataset: only 490, corresponding to 0.172\%. Such a small contamination cannot produce excesses of events detectable by \texttt{RanBox} unless we consider total data samples much larger than a few thousands. In the following we consider data samples of 4900 events in total, in order to allow for up to 10\% signal fractions in our studies.

\begin{table}[h!]
\begin{center}
    \begin{tabular}{rrrrrrrrr}
    \hline
    Test & ${\cal{D}}'$ & $N_s/N_b$ & $R_1$ & $N_{in}$ & $N_{exp}$ & $N_{s}$ & $SR_{1:10}$ & $\overline{\epsilon_s^{1:10}}$\\ % & Vars\\ 
    \hline
    1& 10 & 490/4410 & 268.8& 272 & 0.01 &   0 & 0 & 0.000 \\
    2& 6  & 490/4410 & 215.0& 430 & 1.00 & 430 & 7 & 0.542 \\ % & 0 3 10 11 14 24 \\
    3& 6  & 441/4459 & 247.6& 258 & 0.04 & 258 & 4 & 0.193 \\ % & 0 5 9 10 16 17 \\
    4& 6  & 392/4508 & 214.4& 221 & 0.03 & 221 & 3 & 0.167 \\ % & 2 7 9 11 12 16 \\
    5& 6  & 343/4557 & 217.4& 225 & 0.03 & 225 & 4 & 0.255 \\ % & 0 1 3 11 12 14 \\
    6& 6  & 294/4606 & 202.7& 210 & 0.04 &   0 & 1 & 0.065 \\ % & 4 5 6 24 25 26 \\
    7& 6  & 245/4655 & 202.5& 405 & 1.00 &   0 & 0 & 0.001 \\ % & 3 5 7 8 24 29 \\
\hline
\end{tabular}
\caption{\small Results of \texttt{RanBox} on a 4900-event sample of fraud detection data, with 10,000 trials. The tests have varying signal contamination and subspace dimensionality ${\cal{D}}'$. $SR_{1:10}$ reports the number of signal-rich boxes among the 10 with highest test statistic. See the text for other detail.}  
\label{t:ranbox_fraud}
\end{center}
\end{table}

As shown in Table~\ref{t:ranbox_fraud}, \texttt{RanBox} fails to focus on the fraud component of the data even when its fraction is rather large (10\%), if run with ${\cal{D}}'=10$. However, better results are achieved in this case by reducing the dimensionality of the scanned subspaces: {\em e.g.}, the algorithm shows good sensitivity down to signal fractions of 7\% if the scan is performed in six dimensions.

\begin{table}[h!]
\begin{center}
    \begin{tabular}{rrrrrrrrrr}
\hline
     Test & ${\cal{D}}'_{max}$ & $N_s/N_b$ & $R_1$ & $N_{in}$ & $N_{exp}$ & $N_{s}$ & $SR_{1:10}$ & $\overline{\epsilon_s^{1:10}}$ & ${\cal{D}}'$ \\ % & \\ 
    \hline
    1& 10 & 490/4410 & 307.2 & 321 & 0.04 & 321 &  9 & 0.725 & 9 \\ %& 0 3 10 12 14 21 24    
    2& 10 & 441/4459 & 219.6 & 222 & 0.01 & 222 &  3 & 0.196 & 7 \\ %& 0 3 4 9 10 11 12 14 28
    3& 10 & 392/4508 & 274.5 & 549 & 1.0  &   1 &  0 & 0.001 & 4 \\ %0 12 14 17
    4& 10 & 343/4557 & 263.5 & 275 & 0.04 & 275 & 10 & 0.712 & 7 \\ %0 3 4 6 10 12 14
    5& 10 & 294/4606 & 329.8 & 337 & 0.02 &   0 &  0 & 0.001 & 10\\ % 0 7 10 12 14 16 17 18 26 28  
    6& 10 & 245/4655 & 340.5 & 681 & 1.00 &   1 &  0 & 0.002 & 4 \\ % 0  12 14 21 
    7& 6  & 490/4410 & 222.5 & 445 & 1.00 & 445 & 10 & 0.848 & 6 \\ % 0 3 11 12 14 24
    8& 6  & 441/4459 & 206.0 & 412 & 1.00 & 412 &  3 & 0.266 & 4 \\ % 0 3 14 12
    9& 6  & 392/4508 & 274.0 & 548 & 1.00 &   1 &  3 & 0.189 & 4 \\ % 0 12 14 21
    10&6  & 343/4557 & 249.1 & 254 & 0.02 & 254 &  9 & 0.617 & 6 \\ % 0 3 4 10 12 14 
    11&6  & 294/4606 & 214.0 & 642 & 2.00 &   1 &  0 & 0.001 & 4 \\ % & 0 2 12 14   
    12&6  & 245/4655 & 340.5 & 681 & 1.00 &   1 &  0 & 0.001 & 4 \\ % & 0 3 12 14
    %6  & 196/4704 & 191.0 & 382 & 1.00 &   1 &  0 & 0.005 & 4 \\ % & 4 12 18 26
    \hline
%    480/4320 & 345.3 & 349 & 0.01 & 349 & $\infty$ & \infty & 0.727 \\ % ranboxiter 10 vars alg5
%    384/4416 & 289.87 & 292 & 0.007 & 292 & $\infty$ & $\infty$ & 0.753 \\ % RBI 10 vars alg 5
    
\end{tabular}
\small
\caption{\small Results of \texttt{RanBoxIter} on a 4900-event sample of fraud detection data, using two values for the maximum dimensionality ${\cal{D}}'$ and signal contamination fractions varying from 5\% to 10\%. $SR_{1:10}$ reports the number of signal-rich boxes among the 10 with highest test statistic. See the text for other details.}  
\label{t:ranboxiter_fraud}
\normalsize
\end{center}
\end{table}

\begin{figure}[h!]
\begin{center}
\includegraphics[width=11cm]{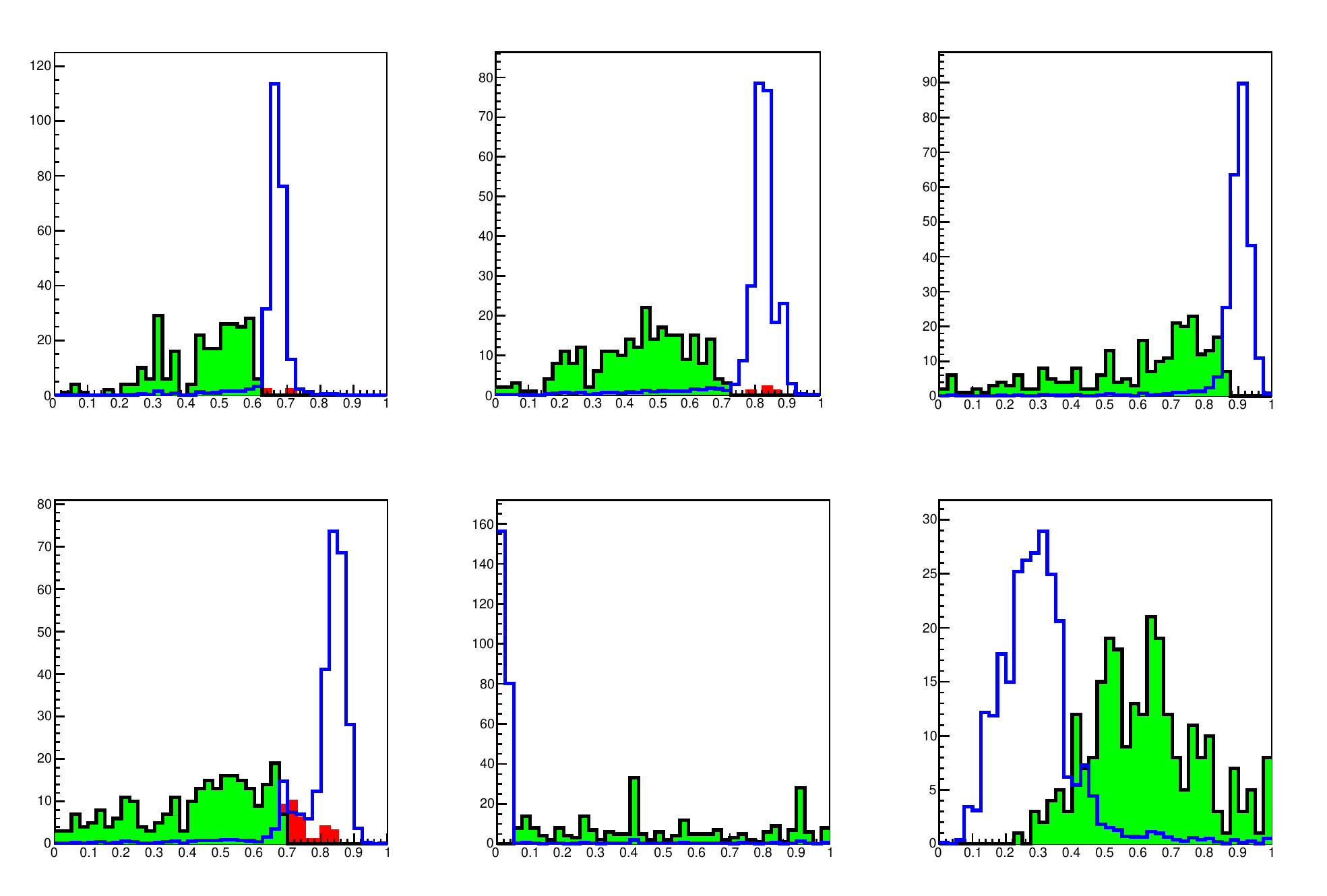}
\includegraphics[width=11cm]{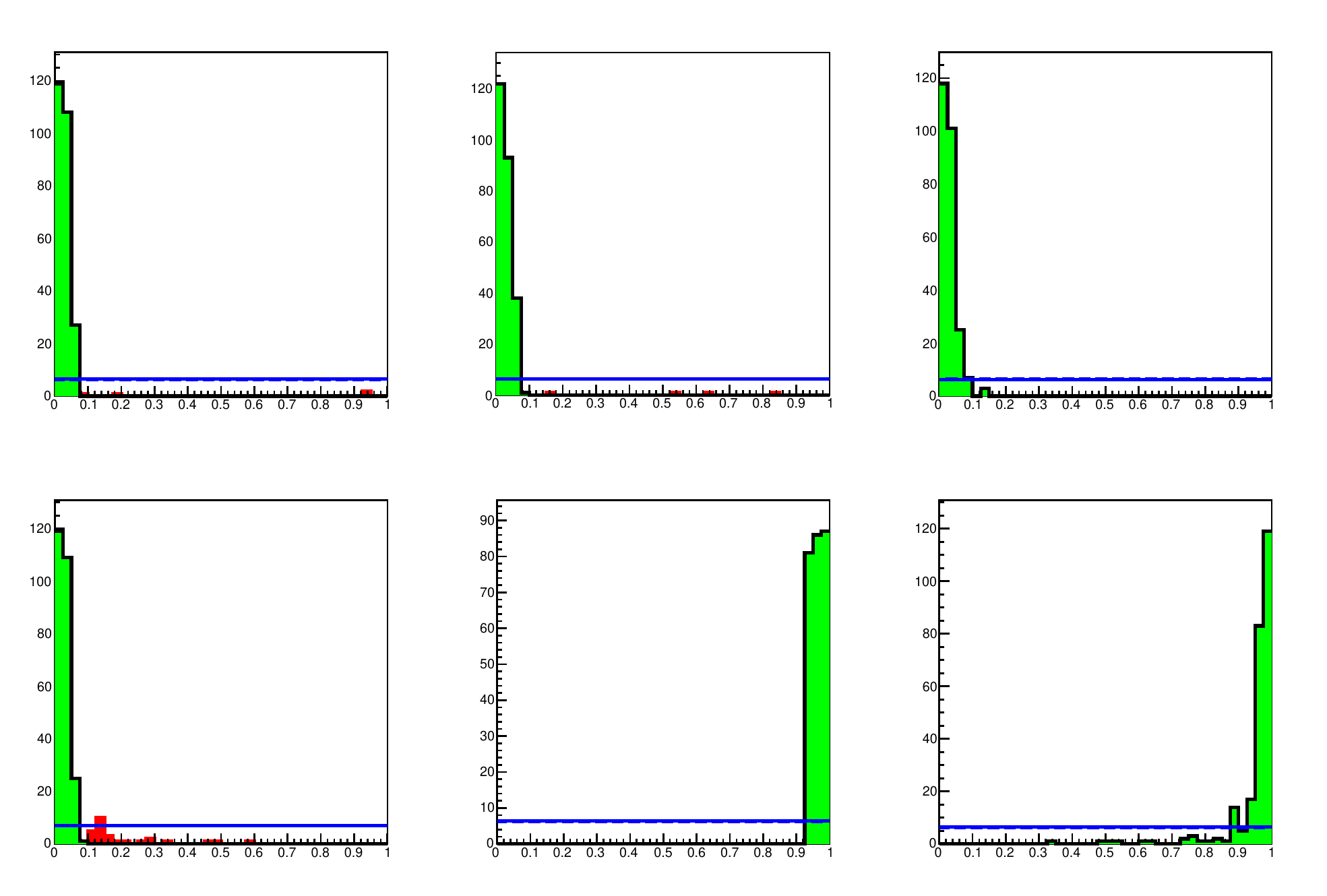}
\small
\caption{\small  Marginal distribution in the original (top two rows) and copula space (bottom two rows) of six features selected by \texttt{RanBoxIter} on a sample of 4900 credit card transactions, with 7\% fraudulent ones (test 10 in Table \ref{t:ranboxiter_fraud}). The six features are the ones defining the subspace selected by the algorithm, where the highest-$R_1$ box contains 100\% of fraudolent transactions. The green distributions refer to the selected events, the blue ones are the original distributions, and red distributions show transactions that do not belong to the box solely because of the value of the displayed feature.}
\normalsize
\label{f:RBI_fraud_uplot}
\end{center}
\end{figure}

\noindent
The iterative version of the algorithm achieves a similar sensitivity to the non-iterative version, as shown in Table~\ref{t:ranboxiter_fraud}. In this case, one notices that it makes little difference to set the maximum dimensionality of the scanned subspaces to the smaller (6) or higher value (10), as in both cases a 7\% signal contamination may be evidenced in the most significant box; Fig.~\ref{f:RBI_fraud_uplot} shows the original and copula space marginals of the features, where test 10 identifies a large signal component. However, one also notices how the results are not always consistent: in the dataset including an 8\% signal contamination (test 3), the scan sticks to a signal-poor region where a background fluctuation creates a locally overdense region in just four dimensions. Rather than investigating further the specific reasons for this behaviour, we catalog it as a further indication that \texttt{RanBox} and \texttt{RanBoxIter} cannot be considered as catch-all anomaly search tools; their power ultimately depends on the specificities of the searched data sample and of the contributing densities, in analogy to the absence of an optimal hypothesis test in the absence of specified alternatives.

\noindent

\clearpage
%%%%%%%%%%%%%%%%%%%%%%%%%%
\section {Related work \label{s:related}}

\noindent
With the exception of a couple of {\em ante litteram} studies that attempted to apply semi-supervised techniques to searches in Tevatron collisions data~\cite{sleuth,d01, d02}, anomaly detection caught the interest of researchers only relatively recently as a viable tool for searches of new phenomena in particle physics. A good review of the status of this area of studies is~\cite{nachman}; a description of some new methods is offered in~\cite{amva4np}. 

In general, all methods that rely on the validity of SM-based simulations for the modeling of backgrounds share a weakness in the large systematic uncertainties of their predictions in those regions of feature space which are the most likely hiding place of new phenomena, and any positive result they yield is liable to raise suspicion ---something that in the past has deterred data analysts from investing efforts in this area of research~\footnote{ The book ``Anomaly! Collider Physics and the Quest of New Phenomena at Fermilab''~\cite{anomaly} describes a number of examples of the general issues connected with the trust of simulations in collaborative searches for new physics signals, taken from the history of the CDF experiment at Fermilab.}. Data-driven approaches are more practicable, at least from a psycho-sociological point of view~\footnote {These issues play a role in collaborative experimentation, as discussed in~\cite{anomaly} as well as in~\cite{staley}.}, but they are usually very specific to the considered signatures (see {\em e.g.}~\cite{resonantnachman} or~\cite{hemispheremixing} for two recent applications), or allow anyway a limited range of uses due to constraining assumptions they rest upon (a good example of this situation is given by the semi-supervised approach pioneered in~\cite{cwola}).

Since an almost ubiquitous feature of new physics signatures in collider searches is constituted by the presence of a bump in an invariant mass distribution, a number of studies (see {\em e.g.}~\cite{nachman1, nachman2} for two recent examples) have concentrated on that distinctive trait in designing algorithms that exploit the signal localization by employing the concept of sidebands: backgrounds usually vary smoothly, so their density can be estimated in the phase space relevant to the signal estimate task by relying on the observed data in regions defined by a positive or negative offset with respect to the assumed mass of the sought resonance. This concept is also used by \texttt{RanBox}, which however extends it to multiple dimensions, moving away from the need to single out one specific variable in the applications. This approach guarantees more generality. 

\noindent
More in general, the task of data-driven anomaly searches in collider data has been addressed with methods that exploit the capability of neural networks or other machine-learning algorithms to provide density estimates~\cite{nndens1,nndens2}, or to map a low-dimensional representation of the data with variational autoencoders~\cite{vae1,vae2,vae3}, or to discriminate a signal by adversarial methods combined with autoencoding~ \cite{vaean1,vaean2,vaegnn}. A large body of relevant literature has built up in the past few years; we direct the interested reader to an up-to-date, living review of such techniques, maintained by Benjamin Nachman in~\cite{livingreview}.

Among all the methods proposed to find new physics in collider data, \texttt{RanBox} shares the most similarities with unsupervised anomaly detection methods that directly attempt a density estimation in the data and construct test statistics to detect anomalous regions, such as ANODE~\cite{nndens1}. Of relevance are also a few other methods that were used in the LHC Olympics anomaly detection challenge~\cite{lhcolympics}, which had the merit to provide a direct comparison of the merits of different approaches to LHC searches. We believe that \texttt{RanBox} however offers some advantages with respect to many of those methods, especially consisting in its lack of strong assumptions about the structure of the data and in its applicability to problems in different domains.  

If we step outside the field of high-energy physics, we find extensive studies of anomaly and outliers detection in statistical and machine learning literature. A number of different approaches have been implemented. Model-based methods~\cite{Barnett} identify a profile of normal instances, then find instances that do not conform to the normal profile, which are considered anomalies; normal profiles can be estimated through classification-based methods~\cite{Abe}, clustering-based methods~\cite{He} or by using robust statistical methods (for a comprehensive monograph on this approach, see~\cite{Rousseeuw1} or~\cite{Rousseeuw3}, and~\cite{Rousseeuw2} for a multivariate example). Other methods are based on distances~\cite{Knorr} and on densities~\cite{Breunig,Liu}. Frauds, which we have considered as a use case in Sec.~\ref{s:experiments} above, are a particularly relevant topic in anomaly detection; widely used tools for fraud detection, most of which are based on unsupervised anomaly identification, are reviewed in~\cite{Bolton}.

\clearpage
%%%%%%%%%%%%%%%%%%%%%%%%%%%%%%%%%%%%%%%%%%%%%%%%%%%%%%%%%%
\section{Concluding remarks \label{s:concludingremarks}}

%Finding structure in multi-dimensional spaces is a task which we can only carry out with automated means, as besides our visualization limits,  we suffer a lack of ability to extrapolate a continuous model out of sparse data. 

\noindent
A number of problems in data science are solved by casting them as ones of supervised learning. Thus, for example, the existence of new elementary particles or phenomena can be routinely tested and excluded in searches at particle colliders by employing simulations of the considered signals as well as of the contributing backgrounds, and classification algorithms that can enrich the data in the signal component. Such a {\em modus operandi} however precludes from our view effects and phenomena for which we have not built, or even considered, a model. To be more inclusive one may cast the problem in the form of anomaly detection: at the expense of renouncing to prior information, we then obtain a more unbiased view of our datasets. In this work we have described and tested a new algorithm which is appropriate for such unbiased searches. By modifying the metric of the space and casting the data in a unit hypercube, a search for regions of the multi-dimensional feature space which contain localized increases in data density becomes feasible even when the original data displays large density variations. The technique has been designed to be applicable to new physics searches, and indeed it proves effective in that domain. However, it also offers itself as a solution to a wide range of problems in unrelated research or non-research fields.

\texttt{RanBox} and \texttt{RanBoxIter} are potentially quite useful tools for the exploration of feature spaces with a dimensionality in the several tens range. We have observed that on particle physics use cases their results compare favourably to those of supervised classification tools such as neural networks or boosted decision trees. For this reason, they should be considered seriously as an investigation tool in collider searches for new phenomena, where each event is comprised typically by few tens of observable features. For still larger dimensions CPU requirements become prohibitive, unless suitable reduction techniques are applied. 

%%%%%%%%%%%%%%%%%%%%%%%%%%%
% \section*{Acknowledgements}

\clearpage
%%%%%%%%%%%%%%%%%%%%%


\begin{thebibliography}{00}

\bibitem{merriam-webster} https://www.merriam-webster.com/dictionary/anomaly.

\bibitem{CMS} The CMS Collaboration, {\em The CMS Experiment at the CERN LHC}, Journ.Inst. 3 (2008) S08004, doi:10.1088/1748-0221/3/08/S08004.

\bibitem{ATLAS} The ATLAS Collaboration, {\em The ATLAS Experiment at the CERN Large Hadron Collider}, Journ. Inst. 3 (2008) S08003, doi:10.1088/1748-0221/3/08/S08003.

\bibitem{gsw} S.L. Glashow, {\em Partial-symmetries of weak interactions}, Nucl. Phys. 22 (1961) 579, doi:10.1016/0029-5582(61)90469-2; S. Weinberg, A Model of Leptons, Phys. Rev. Lett. 19 (1967) 1264, doi:10.1103/PhysRevLett.19.1264; A. Salam, Weak and electromagnetic interactions, in Elementary Particle Physics: relativistic groups and analyticity, N. Svartholm, ed., p.367. Almqvist \& Wiskell, 1968. Proceedings of the 8th Nobel symposium.

\bibitem{ewwg} ALEPH, CDF, D0, DELPHI, L3, OPAL, SLD Collaborations, the LEP Electroweak Working Group, the Tevatron Electroweak Working Group, and the SLD Electroweak and Heavy Flavour Groups, {\em Precision Electroweak Measurements and Constraints on the Standard Model}, CERN PH-EP-2010-095 (2010).


\bibitem{cmsdielectrons} CMS Collaboration, {\em Search for contact interactions and large extra dimensions in the dilepton mass spectra from proton-proton collisions at $\sqrt{s}= 13$ TeV}, JHEP 04 (2019) 114, 	arXiv:1812.10443 [hep-ex], doi:10.1007/JHEP04(2019)114.


\bibitem{sklar} A. Sklar, {\em Fonctions de répartition à n dimensions et leurs marges}, Publ. Inst. Statist. Univ. Paris, 8 (1959) 229.

\bibitem{bellman} R.E. Bellman, Rand Corporation {\em Dynamic programming}, Princeton University Press (1957), p. ix. ISBN 978-0-691-07951-6.

\bibitem{liandma} T.P. Li and Y.Q. Ma, {\em Analysis methods for results in gamma-ray astronomy}, Astroph. Journ. 272 (1983) 317, doi:10.1086/161295.

\bibitem{bonferroni} C.E. Bonferroni, {\em Teoria statistica delle classi e calcolo delle probabilit\`a}, Pubblicazioni del Regio Istituto Superiore di Scienze Economiche e Commerciali di Firenze, 1936.

\bibitem{root} https://root.cern.ch.

\bibitem{cholesky} W.H. Press, S.A. Teukolsky, W.T. Vetterling, and B.P. Flannery, {\em Numerical Recipes in C: The Art of Scientific Computing} (second ed.) (1992). Cambridge University Press, ISBN 0-521-43108-5.

\bibitem{uci} https://archive.ics.uci.edu/ml/index.php.

\bibitem{whiteson} P. Baldi, P. Sadowski, and D. Whiteson, {\em Searching for Exotic Particles in High-Energy Physics with Deep Learning}, Nature Comm. 5 (2014) 4308, doi:10.1038/ncomms5308.

\bibitem{whiteson2} P. Baldi, K. Cranmer, T. Faucett, P. Sadowski, and D. Whiteson, {\em Parameterized Machine Learning for High-Energy Physics}, Eur. Phys. Journ. C76,5 (2016) 7, doi:10.1140/epjc/s10052-016-4099-4.

\bibitem{minibooneUCI} https://archive.ics.uci.edu/ml/datasets/MiniBooNE+particle+identification. 

\bibitem{miniboone} A. A. Aguilar-Arevalo et al., {\em Search for Electron Neutrino Appearance at the $\Delta m^2 \simeq 1 eV^2$ Scale}, Phys. Rev. Lett. 98, 231801 (2007), doi:10.1103/PhysRevLett.98.231801.

\bibitem{lsnd} C. Athanassopoulos et al., {\em Candidate Events in a Search for $\nu \mu \to \nu e$ Oscillations}, Phys. Rev. Lett. 75 (1995) 2650, doi:10.1103/PhysRevLett.75.2650.

\bibitem{roe} H. J. Yang, B. P. Roe, and J. Zhu, "Studies of Boosted Decision Trees for MiniBooNE Particle Identification", Nucl. Instrum. Meth. A555 (2005) 370-385, arXiv:physics/0508045v1 [physics.data-an] (2005), doi:10.1016/j.nima.2005.09.022.

\bibitem{roe2} B. P. Roe, H. J. Yang, J. Zhu, Y. Liu, I. Stancu, and G. McGregor, {\em Boosted Decision Trees as an Alternative to Artificial Neural Networks for Particle
Identification}, Nucl. Instrum. Meth. A543 (2005) 577, arXiv:physics/0408124 [physics.data-an] (2004), doi:10.1016/j.nima.2004.12.018.

\bibitem{creditcarddata} https://www.kaggle.com/mlg-ulb/creditcardfraud.

%%%%%%%%%%%%%%%%%%%%%%%%%%% "related work" references


\bibitem{sleuth} The DZERO Collaboration, {\em A Quasi-Model-Independent Search for New High p\_T Physics at DZero}, Phys. Rev. Lett. 86 (2001) 3712-3717, doi:10.1103/PhysRevLett.86.3712.

\bibitem{d01} D0 collaboration, B. Abbott et al., {\em Search for new physics in $e \mu X$ data at D0 using Sherlock: A quasi model independent search strategy for new physics}, Phys. Rev. D 62
(2000) 092004, [hep-ex/0006011].

\bibitem{d02} D0 collaboration, V. M. Abazov et al., {\em A Quasi model independent search for new physics at large transverse momentum}, Phys. Rev. D 64 (2001) 012004, [hep-ex/0011067].

\bibitem {nachman} B. Nachman, {\em Anomaly Detection for Physics Analysis and Less than Supervised Learning}, arXiv:2010.14554 [hep-ph].

\bibitem {amva4np} A. Stakia et al., {\em Advanced Multi-Variate Analysis Methods for New Physics Searches at the Large Hadron Collider}, arXiv:2105.07530 [hep-ex], submitted to Reviews in Physics.

\bibitem {resonantnachman} J.H. Collins, K. Howe, and B. Nachman, {\em Anomaly Detection for Resonant New Physics with Machine Learning},  Phys. Rev. Lett. 121,  (2018) 241803.

\bibitem {hemispheremixing} P. De Castro Manzano et al., {\em Hemisphere Mixing: a Fully Data-Driven Model of QCD Multijet Backgrounds for LHC Searches}, 	arXiv:1712.02538 [hep-ex], proceedings of the EPS conference, Venice 2017.

\bibitem{anomaly} T. Dorigo, {\em Anomaly! Collider Physics and the Quest for New Phenomena at Fermilab}, World Scientific 2016, doi:10.1142/q0032.

\bibitem{staley} K. Staley, {\em The evidence for the top quark: objectivity and bias in collaborative experimentation}, Cambridge Univ. Press 2004, ISBN: 9780521827102.

\bibitem{cwola} E.M. Metodiev, B. Nachman, and J.Thaler, {\em Classification without labels: Learning from mixed samples in high energy physics}, JHEP 10 (2017) 174, 	arXiv:1708.02949 [hep-ph], doi:10.1007/JHEP10(2017)174.

\bibitem {nachman1} J.H. Collins, K. Howe, B. Nachman, {\em Anomaly Detection for Resonant New Physics with Machine Learning}, Phys. Rev. Lett. 121 (2018) 241803, 	arXiv:1805.02664[hep-ph], doi:10.1103/PhysRevLett.121.241803.

\bibitem {nachman2} J.H. Collins, K. Howe, B. Nachman, {\em Extending the search for new resonances with Machine Learning}, Phys. Rev. D 99, (2019) 014038, arXiv:1902.02634[hep-ph], doi:10.1103/PhysRevD.99.014038.

\bibitem{nndens1} B. Nachman and D. Shih, {\em Anomaly Detection with Density Estimation}, Phys. Rev. D 101 (2020) 075042, 	arXiv:2001.04990 [hep-ph],	doi:10.1103/PhysRevD.101.075042.

\bibitem{nndens2} R.T. D'Agnolo and A. Wulzer, {\em Learning New Physics from a Machine},  Phys. Rev. D 99 (2019) 015014, arXiv:1806.02350 [hep-ph], doi:10.1103/PhysRevD.99.015014.
	

\bibitem {vae1} M. Farina, Y. Nakai, and D. Shih, {\em Searching for New Physics with Deep Autoencoders}, Phys. Rev. D 101 (2020) 075021, 	arXiv:1808.08992 [hep-ph], doi:10.1103/PhysRevD.101.075021.

\bibitem{vae2}  T. Heimel, G. Kasieczka, T. Plehn, J.M. Thompson, {\em QCD or What?}, SciPost Phys. 6 (2019) 030, 	arXiv:1808.08979 [hep-ph], doi:10.21468/SciPostPhys.6.3.030.

\bibitem{vae3} T.S. Roy and A.H. Vijay, {\em A robust anomaly finder based on autoencoders}, 	arXiv:1903.02032 [hep-ph].

\bibitem{vaean1} A. Blance, M. Spannowsky, and P. Waite, {\em Adversarially-trained autoencoders for robust unsupervised new physics searches}, J. High Energy Phys. 2019 (2019) 47, 	arXiv:1905.10384 [hep-ph], doi:10.1007/JHEP10(2019)047.

\bibitem{vaean2} O. Knapp {\em et al.}, {\em Adversarially Learned Anomaly Detection on CMS Open Data: re-discovering the top quark}, arXiv:2005.01598 [hep-ex] (2020).

\bibitem{vaegnn} O. Atkinson, A. Bhardwaj, C. Englert, V.S. Ngairangbam, and M. Spannowsky, {\em Anomaly detection with Convolutional Graph Neural Networks}, arXiv:2105.07988 [hep-ph] (2021).

\bibitem{livingreview} B. Nachman, https://github.com/iml-wg/HEPML-LivingReview.

\bibitem{lhcolympics} G. Kasieczka (ed.) et al., {\em The LHC Olympics 2020: A Community Challenge for Anomaly Detection in High Energy Physics}, arXiv:2101.08320 [hep-ph].

\bibitem {Abe} N. Abe, B. Zadrozny, and J. Langford, {\em Outlier detection by active learning}, in Proceedings of the 12th ACM SIGKDD international conference on Knowledge discovery and data mining, pages 504–509. ACM Press, 2006.

\bibitem{Bay} S. D. Bay and M. Schwabacher, {\em Mining distance-based outliers in near linear time with randomization and a simple pruning rule}, in Proceedings of the ninth ACM SIGKDD international conference on Knowledge discovery and data mining, pages 29–38. ACM Press, 2003.

\bibitem{Barnett} V. Barnett and T. Lewis, {\em Outliers in Statistical Data}, John Wiley \& Sons, Chichester, 1994.

\bibitem{Bolton} R.J. Bolton and D.J. Hand, {\em Statistical Fraud Detection: A Review}, Statist. Sci. 17 (3) 235 - 255, August 2002. https://doi.org/10.1214/ss/1042727940 .

\bibitem {Breunig} M.M. Breunig, H.-P. Kriegel, R.T. Ng, and J. Sander, {\em LOF: identifying density-based local outliers}, ACM SIGMOD Record, 29(2):93–104, 2000.

\bibitem{He} Z. He, X. Xu, and S. Deng, {\em Discovering cluster-based local outliers}, Pattern Recogn. Lett., 24(9-10):1641–1650, 2003.

\bibitem {Knorr} E.M. Knorr and R.T. Ng, {\em Algorithms for mining distancebased outliers in large datasets}, in VLDB ’98: Proceedings of the 24rd International Conference on Very Large Data Bases, pages 392–403, San Francisco, CA, USA, 1998. Morgan Kaufmann.

\bibitem{Liu} F.T. Liu, K.M. Ting, and Z. Zhou, {\em Isolation Forest}, 2008 Eighth IEEE International Conference on Data Mining, 2008, pp. 413-422, doi:10.1109/ICDM.2008.17.

\bibitem{Rousseeuw1} P.J. Rousseeuw and M.Hubert, {\em Anomaly detection by robust statistics}, WIREs Data Mining Knowl. Discov. 2018, 8:e1236. doi:10.1002/widm.1236.

\bibitem{Rousseeuw3} P.J. Rousseeuw and A.M. Leroy, {\em Robust Regression and Outlier Detection}, New York: Wiley-Interscience; 1987.   

\bibitem{Rousseeuw2} P.J. Rousseeuw and K.V. Driessen, {\em A fast algorithm for the minimum covariance determinant estimator}, Technometrics, 41(3):212–223, 1999.


%%%%%%%%%%%%%%%%%%%%%
\end{thebibliography}
\end{document}